%% file: paper.tex
\documentclass{aa}  

\usepackage{graphicx}
\usepackage{lscape}
\usepackage{natbib} 
\usepackage{siunitx}
\usepackage{url}

\usepackage[varg]{txfonts}

\bibpunct{(}{)}{;}{a}{}{,} 

\begin{document}

\title{PoWR grids of non-LTE model atmospheres for OB-type stars of various metallicities}

   \author{R. Hainich\inst{1}
          \and V. Ramachandran\inst{1}
          \and T. Shenar\inst{1,2}
          \and A.\,A.\,C. Sander\inst{1,3}
          \and H. Todt\inst{1}
          \and D. Gruner\inst{1}
          \and L. M. Oskinova\inst{1}
          \and W.-R. Hamann\inst{1}
          }

   \institute{Institut f\"ur Physik und Astronomie,
              Universit\"at Potsdam,
              Karl-Liebknecht-Str. 24/25, D-14476 Potsdam, Germany \\
              \email{rhainich@astro.physik.uni-potsdam.de}
              \and
              Institute of astrophysics, KU Leuven, 
              Celestijnlaan 200D, 3001 Leuven, Belgium
              \and
              Armagh Observatory and Planetarium, 
              College Hill, Armagh, BT61 9DG, Northern Ireland
              }
   \date{Received <date> / Accepted <date>}


\abstract{
The study of massive stars in different metallicity environments is a central topic of current stellar research. The spectral analysis of massive stars requires adequate model atmospheres. 
The computation of such models is difficult and time-consuming. Therefore, spectral analyses are greatly facilitated if they can refer to existing grids of models. Here we provide grids of 
model atmospheres for OB-type stars at metallicities corresponding to the Small and Large Magellanic Clouds, as well as to solar metallicity. In total, the grids comprise 785 individual models.
The models were calculated using the state-of-the-art Potsdam Wolf-Rayet (PoWR) model atmosphere code. The parameter domain of the grids was set up using stellar evolution tracks. For all these models, we provide normalized and flux-calibrated spectra, spectral energy distributions, feedback parameters such as ionizing photons, Zanstra temperatures, and photometric magnitudes. The atmospheric structures (the density and temperature stratification) are available as well. All these data are publicly accessible through the PoWR website.
}
\keywords{Stars: massive -- Stars: early type -- Stars: atmospheres -- Stars: winds, outflows -- Stars: mass-loss -- Radiative transfer}

\maketitle

\section{Introduction}
\label{sect:intro}

Through their powerful stellar winds, ionizing fluxes, 
and supernova (SN) explosions, massive stars ($M_\text{i} \gtrsim 8\,M_\odot$) dominate the energy budget of their host galaxies. They are the progenitors of core-collapse SNe, leaving behind a neutron star (NS) or a black hole (BH), which makes them central players in modern gravitational-wave (GW) astrophysics 
\citep[e.g.,][]{Marchant2016, deMink2016, Hainich2018}. 
Spectroscopically, they are predominantly identified with O and early B spectral types. When surrounded by thick stellar winds, 
they are classified as Wolf-Rayet (WR) stars \citep{Smith1968, Smith1996}, as transition-type stars, such as Of/WN stars \citep[e.g.,][]{Crowther2011}, or as luminous blue variables \citep[LBVs; e.g.,][]{Humphreys1994,vanGenderen2001} 

In recent years, the topic of massive stars at low metallicity $(Z)$ has been gaining tremendous momentum. The first stars that formed 
in our universe must have been born in extremely $Z$-poor environments \citep{Bromm2004}. Massive binaries at low $Z$ are the leading candidates for massive GW merger systems \citep[e.g.,][]{Eldridge2016}. Generally, 
massive stars as a function of $Z$ are intensively studied; for example, the $Z-$dependence of multiplicity parameters 
\citep{Sana2013, Almeida2017}, initial masses \citep{Schneider2018}, 
binary interaction physics \citep{Foellmi2003a, Shenar2016, Shenar2017}, stellar feedback \citep{Ramachandran2018,Ramachandran2018b}, stellar rotation \citep{Meynet2005}, and stellar winds \citep{Mokiem2007, Hainich2015}. The Small and Large Magellanic Clouds (SMC, LMC), with their well-constrained distances, low interstellar extinctions, and subsolar metallicity of ${\sim}1/7$ and $1/2$ solar, respectively \citep{Dufour1982, Larsen2000, Trundle2007}, offer ideal laboratories for studying $Z$-dependent effects.

The physical parameters of massive stars, such as their temperatures, luminosities, and masses, can be derived by comparing observed to synthetic spectra.
To model massive star atmospheres, 
it is essential to allow for non-local thermodynamic equilibrium (non-LTE), 
and to account for the millions of iron-line transitions in the ultraviolet (UV) that give rise to the so-called line-blanketing \citep[e.g.,][]{Hubeny1995, Hillier1998}. For most O-type stars, as well as for evolved B-type stars, a calculation of the wind is also required \citep{Hamann1981, Kudritzki1992, Puls2008}. There are only a few codes worldwide that fulfill these requirements \citep[see overviews in, e.g.,][]{Puls2008b, Sander2015}.

\begin{figure*}[tbp]
    \centering
    \includegraphics[angle=-90,width=0.8\hsize]{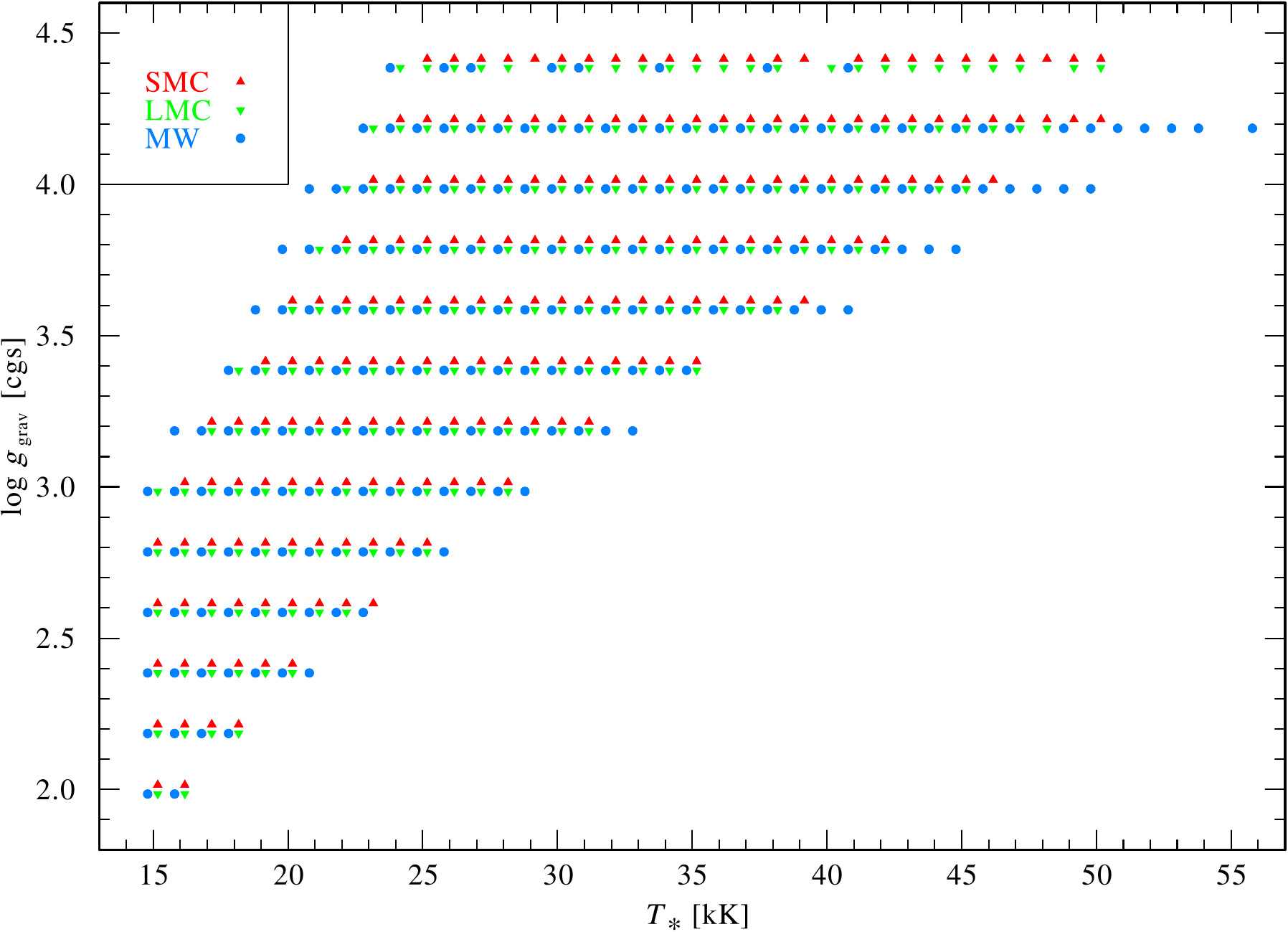}
    \caption{
        Overview of OB-type model grids in the $T_\ast - \log g_\mathrm{grav}$ plane. Each symbol represents an available PoWR model. The different colors and symbols indicate the different grids described in Sect.\,\ref{sect:grids}. The extension of the two SMC grids is identical.
    }
    \label{fig:gitter}
\end{figure*}

The Potsdam Wolf-Rayet (PoWR) model atmosphere program is one of these codes. Originally developed for WR stars, 
it is now applicable to any hot star that does not show significant deviations from spherical symmetry,
including OB-type stars \citep{Graefener2002, Hamann2003, Sander2015}. Using PoWR, fundamental parameters have been derived for many WR stars and binaries in the Galaxy \citep{Hamann1995, Sander2012} and the Magellanic Clouds \citep{Hainich2014, Shenar2016}, as well as for OB-type stars and binaries \citep{Shenar2015, Ramachandran2018}. PoWR model grids for WR stars of various types and at various metallicites have been published online\footnote{\url{www.astro.physik.uni-potsdam.de/PoWR}\label{fnote}} \citep{Sander2012, Hamann2004, Todt2015}. With the current paper, we announce the publication of extensive model grids of OB-type stars at SMC, LMC, and solar metallicities calculated with the PoWR code. The applicability of these model grids ranges from spectral analyses of OB-type stars to theoretical applications that need model spectra as an input such as population synthesis \citep[e.g.,][]{Leitherer2014,Eldridge2017}, or applications that require atomic level population numbers as input such as three-dimensional (3D) Monte-Carlo calculations of stellar winds \citep[e.g..][]{Surlan2012a,Surlan2012b}.

The paper is structured as follows. In Sect.\,\ref{sect:models} we describe the basics of the PoWR atmosphere models. The OB-type grids, the data products, and the web interface are introduced in Sect.\,\ref{sect:grids}. In Sect.\,\ref{sect:conclusions} we discuss some findings based on our model calculations. Finally, we give a short overview of potential applications in Sect.\,\ref{sect:summary}.

\section{The models}
\label{sect:models}

The synthetic spectra presented in this work are calculated with the Potsdam Wolf-Rayet (PoWR) code, which is a state-of-the-art code for expanding stellar atmospheres. PoWR assumes spherical symmetry and   
a stationary outflow.
It accounts for non-LTE effects, a consistent stratification in the hydrostatic (lower) part of the atmosphere, iron line blanketing, and wind inhomogeneities. The code solves the rate equations for the statistical equilibrium simultaneously with the radiative transfer, which is calculated in the comoving frame. At the same time, the code ensures energy conservation. For details on the code, we refer to \citet{Graefener2002}, \citet{Hamann2003}, \citet{Todt2015}, and \citet{Sander2015}.

The main parameters of OB-type models are the stellar temperature $T_\ast$, the luminosity $L$, the surface gravity $\log g_\mathrm{grav}$, the mass-loss rate $\dot{M}$, and the terminal wind velocity $v_\infty$. The stellar temperature and the luminosity specify the stellar radius $R_\ast$ via the Stefan-Boltzmann law 
\begin{equation}
\label{eq:sblaw}
L = 4 \pi \sigma_\mathrm{SB} R_\ast^2 T_\ast^4~.
\end{equation} 
The stellar radius is by definition the inner boundary of the model atmosphere, which we locate at a Rosseland continuum optical depth of $\tau_\mathrm{Ross} = 20$. The stellar temperature $T_*$ is then the effective temperature that corresponds to $R_\ast$. The outer boundary is set to $R_\mathrm{max} = 100\,R_\ast$.

In the subsonic part of the stellar atmosphere, the velocity field $v(r)$ is calculated self-consistently such that a quasi-hydrostatic density stratification is obtained. A classical $\beta$-law \citep{Castor1979,Pauldrach1986}
\begin{equation}
\label{eq:betalaw}
v(r) = v_\infty \left ( 1 - \frac{R_0}{r}\right )^{\beta}
,\end{equation} 
with $R_0 \approx R_\ast$ is assumed in the wind, which corresponds to the supersonic part of the atmosphere. For the exponent, the value $\beta = 0.8$ is assumed for all models \citep{Kudritzki1989,Puls1996}. 

In the comoving-frame calculations, turbulent motion is accounted for by using Gaussian line profiles with a Doppler width of $30\,\textrm{km/s}$.
This choice is motivated by the requirement to limit the computation time; tests revealed that narrower line profiles during the comoving-frame calculations have very limited impact on the resulting stratification. In the hydrostatic equation, the turbulent pressure is taken into account by means of a microturbulent velocity $\xi$ \citep[see ][]{Sander2015}.

After an atmosphere model is converged, the synthetic spectrum, also denoted as emergent spectrum, is calculated by integrating the source function in the observers frame along emerging rays parallel to the line-of-sight.
In this formal integral the Doppler velocity is decomposed
into a depth-dependent thermal component and the microturbulent velocity,
which is set to $\xi(R_*) = 14\,\mathrm{km/s}$
at the base of the wind and grows proportional to the wind velocity up to a value of $\xi(R_\mathrm{max}) = 0.1\,v_\infty$. 

Wind inhomogeneities are accounted for by assuming optically thin clumping. The clumping factor $D$ (which is the inverse of the volume filling factor, $f_\mathrm{V} = D^{-1}$) describes the over-density in the clumps compared to a homogeneous model with the same mass-loss rate \citep{Hillier1991,Hamann1998}, while the interclump medium is considered to be void. 
We assume that clumping starts at the sonic point and reaches its maximum value $D=10$ at a stellar radius of $10\,R_\ast$ \citep[cf.][]{Runacres2002}.

Detailed model atoms of \element{H}, \element{He}, \element{C}, \element{N}, \element{O}, \element{Mg}, \element{Si}, \element{P}, and \element{S} were included in the non-LTE calculations (see Table\,\ref{table:model_atoms}). The iron group elements (\element{Sc}, \element{Ti}, \element{V}, \element{Cr}, \element{Mn}, \element{Fe}, \element{Co}, and \element{Ni}) with their multitude of levels and line transitions were treated in a superlevel approach \citep[see\ ][]{Graefener2002}, combining levels and transitions into superlevels with pre-calculated transition cross-sections and with the assumption of solar abundance ratios relative to iron. 

\begin{figure}[tbp]
    \centering
    \includegraphics[width=\hsize]{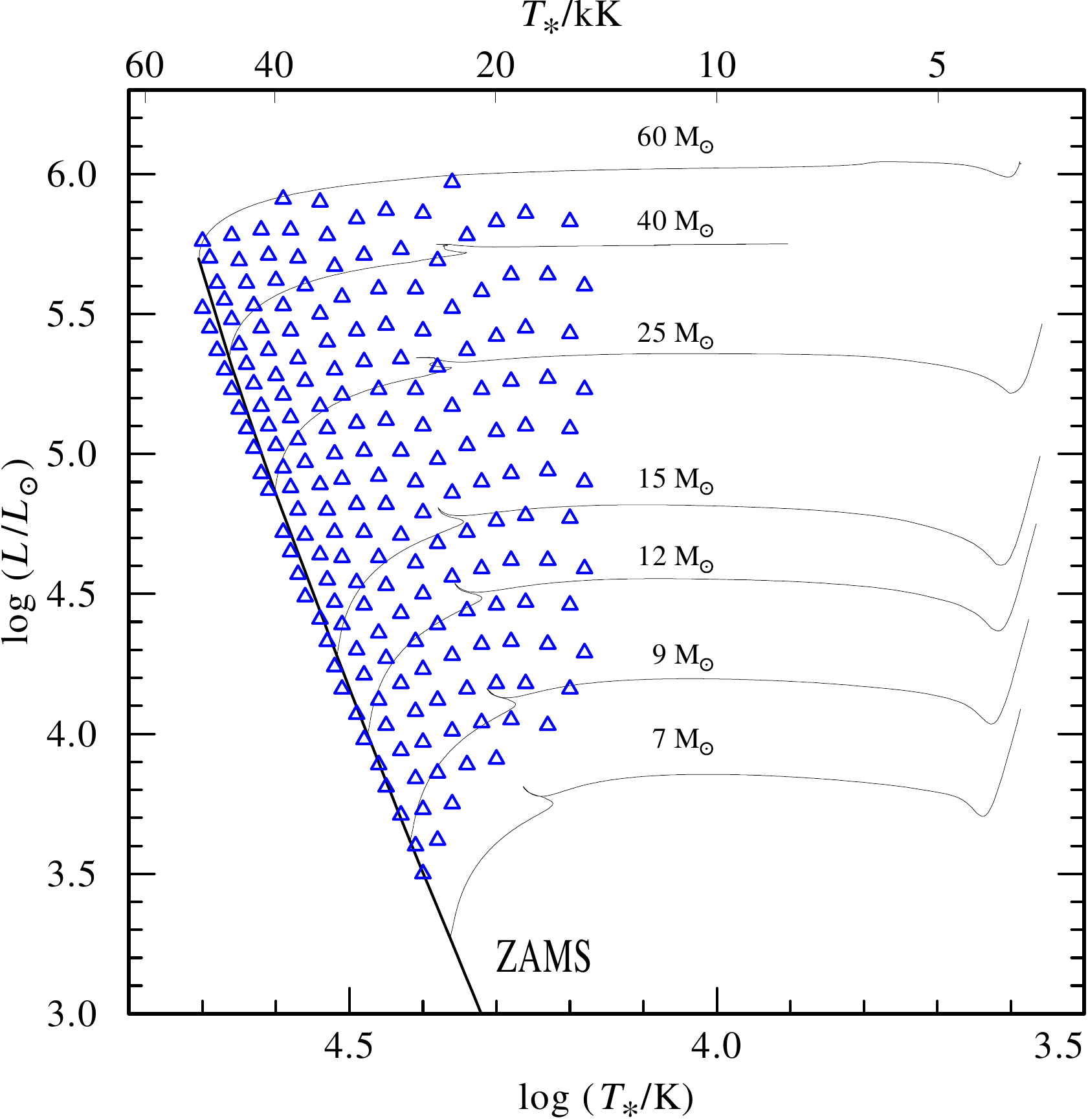}
    \caption{
        Hertzsprung-Russell diagram illustrating the coverage of the $\log T_\ast$-$\log L$ domain by our SMC model-grid. Each blue triangle refers to one grid model. The depicted stellar evolution tracks were calculated by \citet{Brott2011}.
    }
    \label{fig:hrd_smc}
\end{figure}

\begin{figure*}[tbp]
    \centering
    \includegraphics[width=0.95\hsize]{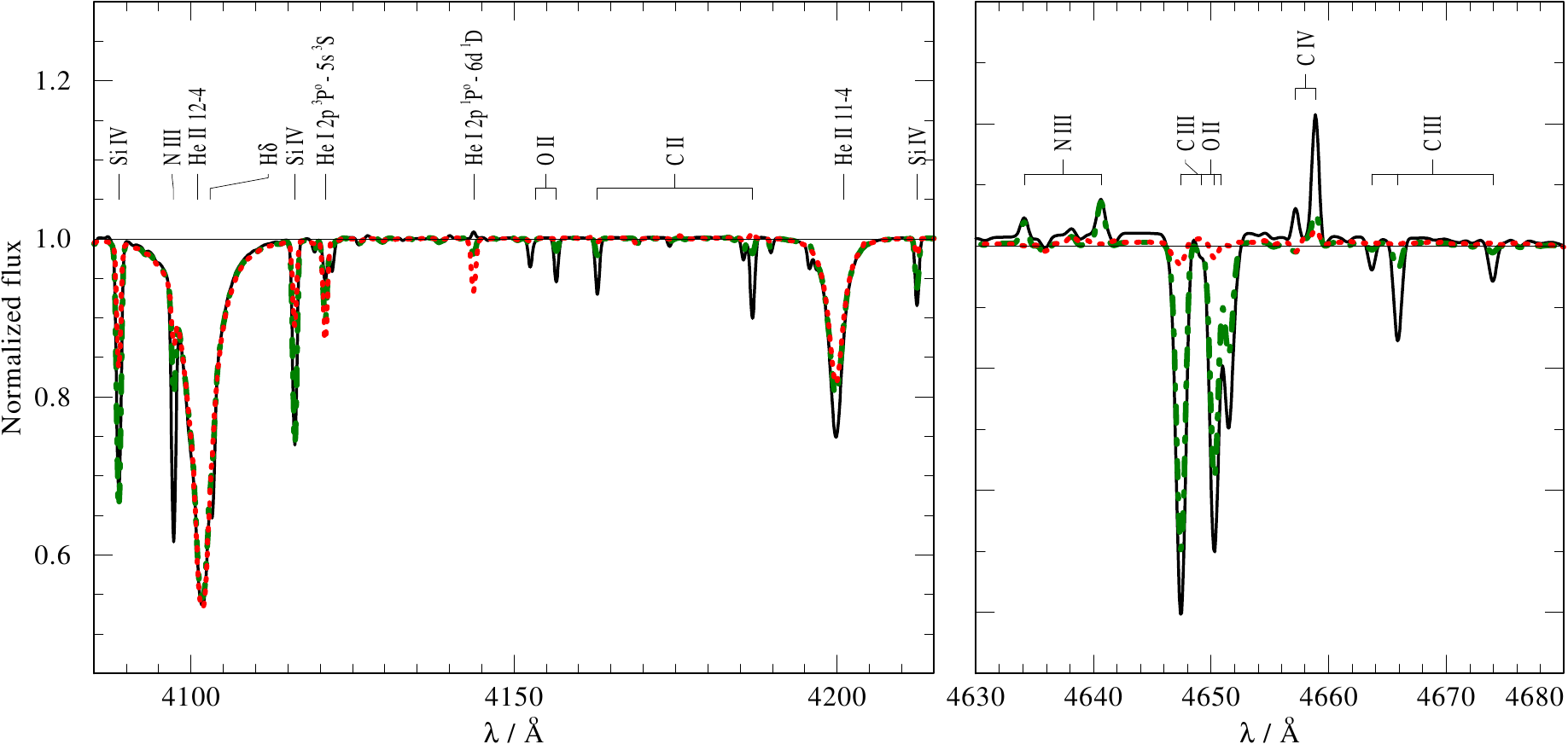}
    \caption{
        Normalized line spectra of the models with $T_\ast = 36\,\mathrm{kK}$ and $\log g_\mathrm{grav} = 3.8\,\mathrm{[cgs]}$ from the grid with SMC (red dotted line), LMC (green dashed line), and solar (black continuous line) metallicity. The mass-loss rate of all models is 
        $\dot{M} = 10^{-7} M_\odot/\mathrm{yr}$. Two exemplary wavelength  ranges with prominent metal lines are depicted.
    }
    \label{fig:spec_comp}
\end{figure*}

\section{The OB-type atmosphere grids}
\label{sect:grids}

The PoWR code is employed to construct grids of model atmospheres for early B-type and O-type stars. Altogether four grids have been calculated so far, ranging from solar metallicities down to LMC ($Z_\textrm{LMC} \approx 1/2\,Z_\odot$) and SMC metallicities ($Z_\textrm{SMC} \approx 1/7\,Z_\odot$, \citealt{Dufour1982,Larsen2000,Trundle2007}). Two grids have been established for the SMC, which correspond to two different mass-loss rates, while only one grid has been calculated for the LMC and one for the Galaxy. The parameterization of the grids is illustrated in Fig.\,\ref{fig:gitter}. In total 785 models have been calculated. Further grids that will improve the coverage of the mass-loss domain are currently in preparation and will be made available in the near future. 

The independent parameters of the grid models are the stellar temperature $T_\ast$ and the surface gravity. The grid spacing is 1\,kK for $T_\ast$ and 0.2\,dex for $\log g_\mathrm{grav}$. The gravitational acceleration is given by
\begin{equation}
\label{eq:ggrav}
g_\mathrm{grav} = \frac{G M}{R_\ast^2},
\end{equation} 
where $M$ is the stellar mass and $G$ the gravitational constant. 
For the SMC and LMC grids, models have been calculated for stellar temperatures of 15\,kK to 50\,kK, while the temperature range is $15 - 56$\,kK for the Galactic grid. 
Besides $T_\ast$ and $\log g_\mathrm{grav}$, the luminosity is a further model parameter.
The value of $L$ has been set 
by using stellar evolution tracks and by interpolating among them. Because of that, the extension of the grids in the $\log g_\mathrm{grav}$ domain is limited by the coverage of the stellar evolution tracks.
This is illustrated in Fig.\,\ref{fig:hrd_smc} that depicts a Hertzsprung-Russell diagram (HRD) with the SMC model grid and the stellar evolution models used to construct this grid. The corresponding plots for the other grids can be found in Appendix\,\ref{sec:addfigures}.
For the SMC and LMC grids, the stellar evolution models calculated by \citet{Brott2011} were employed, while the models by \citet{Ekstroem2012} were used for the Galactic grid, since those evolution models have a superior coverage of the initial mass domain. These different sets of evolution models are the reason why the extension of the grids is not the same for the MW, LMC, and the two SMC grids as visible in Fig.\,\ref{fig:gitter}.

Based on $T_\ast$, $L$, and $\log g_\mathrm{grav}$, 
the escape velocity for each model is calculated, which in turn is used to estimate the terminal wind velocity by applying the scaling relations established by \citet{Lamers1995}. Accounting for the hot bi-stability jump, a factor of 1.3 is used for stars with $T_\ast < 21\,\mathrm{kK}$, while 2.6 is applied above 21\,kK \citep[see also ][]{Lamers1999}. 
In addition, the terminal wind velocities for the SMC and LMC models are scaled with $(Z/Z_\odot)^{0.13}$, following \citet{Leitherer1992}.

A further model parameter is the mass-loss rate $\dot{M}$ or, equivalently, the wind strength parameter $\log\,Q$, which is used instead of $\dot{M}$ in the two SMC grids to prescribe the wind mass-loss. In the PoWR code, the following definition of the $\log Q$ parameter is adopted
\begin{equation}
\label{eq:logQ}
Q = \frac{\dot{M} / ( M_\odot/\mathrm{yr}) \cdot D^{1/2}}{(v_\infty / (\mathrm{km/s}) \cdot R_* / R_\odot)^{3/2}}
,\end{equation} 
\citep[see e.g.,][]{Puls1996,Puls2008,Sander2017a}.
The two SMC grids are calculated with $\log Q = -13.0$ and $\log Q = -12.0$, respectively. The use of a fixed $\log Q$ in those grids implies that the mass-loss rate is not constant throughout the grids, since $v_\infty$ and $R_\ast$ vary from model to model.
In the MW and LMC grid, we instead used a fixed mass-loss rate of $\dot{M} = 10^{-7} M_\odot/\mathrm{yr}$ for all models.    
This value of $\dot{M}$ is chosen because 
our grids are meant as an extension of the parameter space of earlier grids, such as those published by \citet{Lanz2003,Lanz2007}, to significant mass-loss rates. Hence, a certain amount of wind is always present in our models.
The models calculated by \citet{Lanz2003,Lanz2007} with their TLUSTY code adopt the approximation of a plane-parallel and static atmosphere. \citet{Sander2015} showed that in the limit of vanishing $\dot{M}$ and infinite curvature radius, the emergent spectra of PoWR model atmospheres agree very well with the TLUSTY results.

The mass-loss rate is an import parameter that significantly determines the density in the wind and, consequently, also the emergent spectrum. The spectral range that is influenced the most by the choice of $\dot{M}$ is the UV with its key diagnostic wind lines such as \ion{N}{v}\,$\lambda\lambda1239, 1243$\,\AA,  \ion{Si}{iv}\,$\lambda\lambda1393.8, 1402.8$\,\AA, \ion{C}{iv}\,$\lambda\lambda1548, 1550.8$\,\AA, \ion{He}{ii}\,$\lambda1640$\,\AA\ and \ion{N}{iv}\,$\lambda1718$\,\AA. Due to the choice of modest mass-loss rates for the presented grids, the emergent spectra of all models show at least some of those lines in the form of P Cygni profiles, depending on the specific ionization structure. In comparison to the UV, the optical wavelength range is significantly less influenced by mass loss. In this range, the main wind-contaminated lines are \ion{He}{ii}\,$\lambda4686$\,\AA\ and \element{H}$\alpha$. While \element{H}$\alpha$ might show a certain amount of wind emission in its profile for the cool models with low surface gravities, the adopted mass-loss rates are usually too low to push \ion{He}{ii}\,$\lambda4686$\,\AA\ into emission. Besides these two prominent lines, weaker nitrogen and carbon lines might appear in emission, as illustrated in the right panel of Fig.\,\ref{fig:spec_comp}. In the infrared (IR), the most prominent line that is influenced by the wind and consequently by $\dot{M}$ is Br$\gamma$, which shows an emission component preferentially in the O-star models.

For the galactic grid, we assume solar abundances as derived by \citet{Asplund2009}. In the LMC and SMC models, we adopt the abundances obtained by \citet{Hunter2007} and \citet{Trundle2007} for C, N, O, Mg, Si, and Fe.
For P and S, we use the corresponding solar abundances, scaled to the metallicity of the LMC and SMC by a factor of 1/2 and 1/7, respectively. The hydrogen mass fraction is set to $X_\mathrm{H} = 0.74$ in all models. 

In the comoving-frame calculations of the LMC and MW grid models, a micro turbulent velocity of $\xi = 10\,\mathrm{km/s}$ is used, while the SMC models are calculated with $\xi = 14\,\mathrm{km/s}$.

\subsection{Data products}
\label{sect:data}

The most important output of the model calculations are the synthetic line spectra. We provide a continuous coverage from the UV to the near-IR (NIR) ($920\,\AA - 2.4\,\mu\mathrm{m}$), including the K-band, as well as a significant fraction of the mid-IR domain (10 to $20\,\mu\mathrm{m}$). 
These emergent spectra are calculated in the observer's frame and have a spectral resolution of about $R = 160.000$ (corresponding to 5\,km/s in the velocity space).  
Flux-calibrated and continuum-normalized spectra are available. The 
normalized line spectra of two exemplary wavelength ranges with prominent metal lines are displayed in Fig.\,\ref{fig:spec_comp}. This figure illustrates the spectral differences between late O-type giants at different metallicities by comparing models with the same $T_\ast$, $\log g_\mathrm{grav}$, and $\dot{M}$ from the different models grids.

\begin{figure}[tbp]
    \centering
    \includegraphics[width=\hsize]{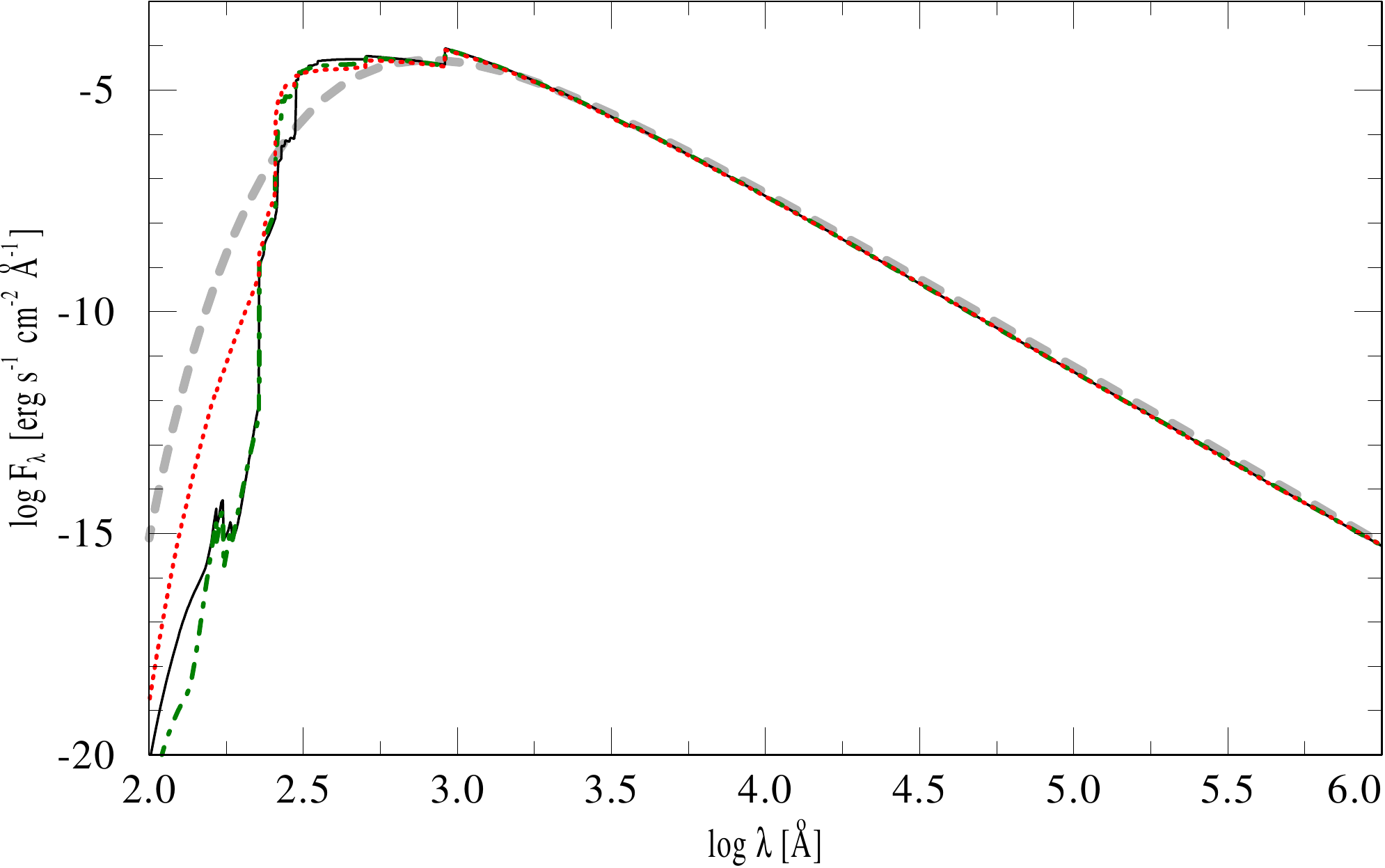}
    \caption{
        Spectral energy distributions of the three models shown in Fig.\,\ref{fig:spec_comp} in comparison with a black body of the same effective temperature as the models. The models are plotted with the same line styles and colors as in Fig.\,\ref{fig:spec_comp}, while the black body is depicted by a thick gray dashed line.
    }
    \label{fig:comp_sed}
\end{figure}

We also provide spectral energy distributions (SEDs) over the whole spectrum.
These SEDs include all lines but are on a coarse wavelength grid and were calculated in the comoving frame. The SEDs for the three models depicted in Fig.\,\ref{fig:spec_comp} are plotted in Fig.\,\ref{fig:comp_sed} in comparison to a back body with $T_\mathrm{eff} = 36\,\mathrm{kK}$.

Feedback parameters such as the number of hydrogen ionizing and helium ionizing photons and Zanstra temperatures are available for all models. In addition, we provide Johnson $U$, $B$, and $V$ magnitudes and Stroemgren $u$, $v$, $b$, and $y$ magnitudes. In Table\,\ref{tab:feedback}, the predicted magnitudes and feedback parameters are listed, exemplary for the models shown in Figs.\,\ref{fig:spec_comp} and \ref{fig:comp_sed}.

\begin{table}
    \caption{Feedback parameters and magnitudes of the three models from Figs.\,\ref{fig:spec_comp}, \ref{fig:comp_sed}, and \ref{fig:comp_He}.}
    \label{tab:feedback}      
    \centering
    \begin{tabular}{lSSS}
        \hline\hline 
        \rule[0mm]{0mm}{2.2ex} 
         & \multicolumn{1}{c}{MW} & \multicolumn{1}{c}{LMC} & \multicolumn{1}{c}{SMC} \\
        \hline 
        $\log Q_\mathrm{\element{H}}\, [\mathrm{s^{-1}}]$    & 48.83 & 48.82 & 48.82 \rule[0mm]{0mm}{2.5ex} \\
        $T_{\mathrm{Zanstra}, \element{H}}\, [\mathrm{kK}]$  & 35.5  & 35.3  & 35.4  \\
        $\log Q_\mathrm{\ion{He}{i}}\, [\mathrm{s^{-1}}]$    & 47.80 & 47.78 & 47.81 \\
        $\log Q_\mathrm{\ion{He}{ii}}\, [\mathrm{s^{-1}}]$   & \multicolumn{1}{c}{-\tablefootmark{b}} & \multicolumn{1}{c}{-\tablefootmark{b}} & 42.06 \\
        $T_{\mathrm{Zanstra}, \element{He}}\, [\mathrm{kK}]$ & \multicolumn{1}{c}{-\tablefootmark{b}} & \multicolumn{1}{c}{-\tablefootmark{b}} & 27.9  \\
        $M_\mathrm{U}\, [\mathrm{mag}]$                      & -6.60 & -6.58 & -6.54 \\
        $M_\mathrm{B}\, [\mathrm{mag}]$                      & -5.42 & -5.41 & -5.37 \\
        $M_\mathrm{V}\, [\mathrm{mag}]$                      & -5.12 & -5.11 & -5.08 \\
        $M_\mathrm{u}\, [\mathrm{mag}]$\tablefootmark{a}     & -5.52 & -5.50 & -5.46 \\
        $M_\mathrm{b}\, [\mathrm{mag}]$\tablefootmark{a}     & -5.23 & -5.23 & -5.19 \\
        $M_\mathrm{v}\, [\mathrm{mag}]$\tablefootmark{a}     & -5.31 & -5.31 & -5.27 \\
        $M_\mathrm{y}\, [\mathrm{mag}]$\tablefootmark{a}     & -5.14 & -5.14 & -5.10 \\
        \hline 
    \end{tabular}
    \tablefoot{
        \tablefoottext{a}{Stroemgren magnitudes}
        \tablefoottext{b}{For these relatively cool models, the \ion{He{ii}} ionizing flux ($\lambda < 228$\,\AA) is neglidgeble.}
    }
\end{table}

The atmospheric structure (e.g., the density and the velocity stratification) is supplied for all models. As an example, the structure information of the model with $T_\ast = 25\,\mathrm{kK}$ and $\log g_\mathrm{grav} = 3.2\,\mathrm{[cgs]}$ from the LMC model grid is listed in Table\,\ref{table:structure}.

\subsection{The web interface}

All information described in Sect.\,\ref{sect:data} can be accessed via the PoWR web interface\footnote{\url{www.astro.physik.uni-potsdam.de/PoWR}}. A general description of the interface and how to use it can be found in \citet{Todt2015}. Recently, an option to obtain the tabulated atmospheric structure and the possibility to download the selected data product for a whole grid was added to the online interface. Both these options are available after having selected a specific model from the grids.
More detailed information such as the population numbers or high-resolution SEDs calculated in the observer's frame are currently not accessible via the web interface, but can be provided on individual request.

\section{Discussion}
\label{sect:conclusions}

\begin{figure}[tbp]
    \centering
    \includegraphics[width=\hsize]{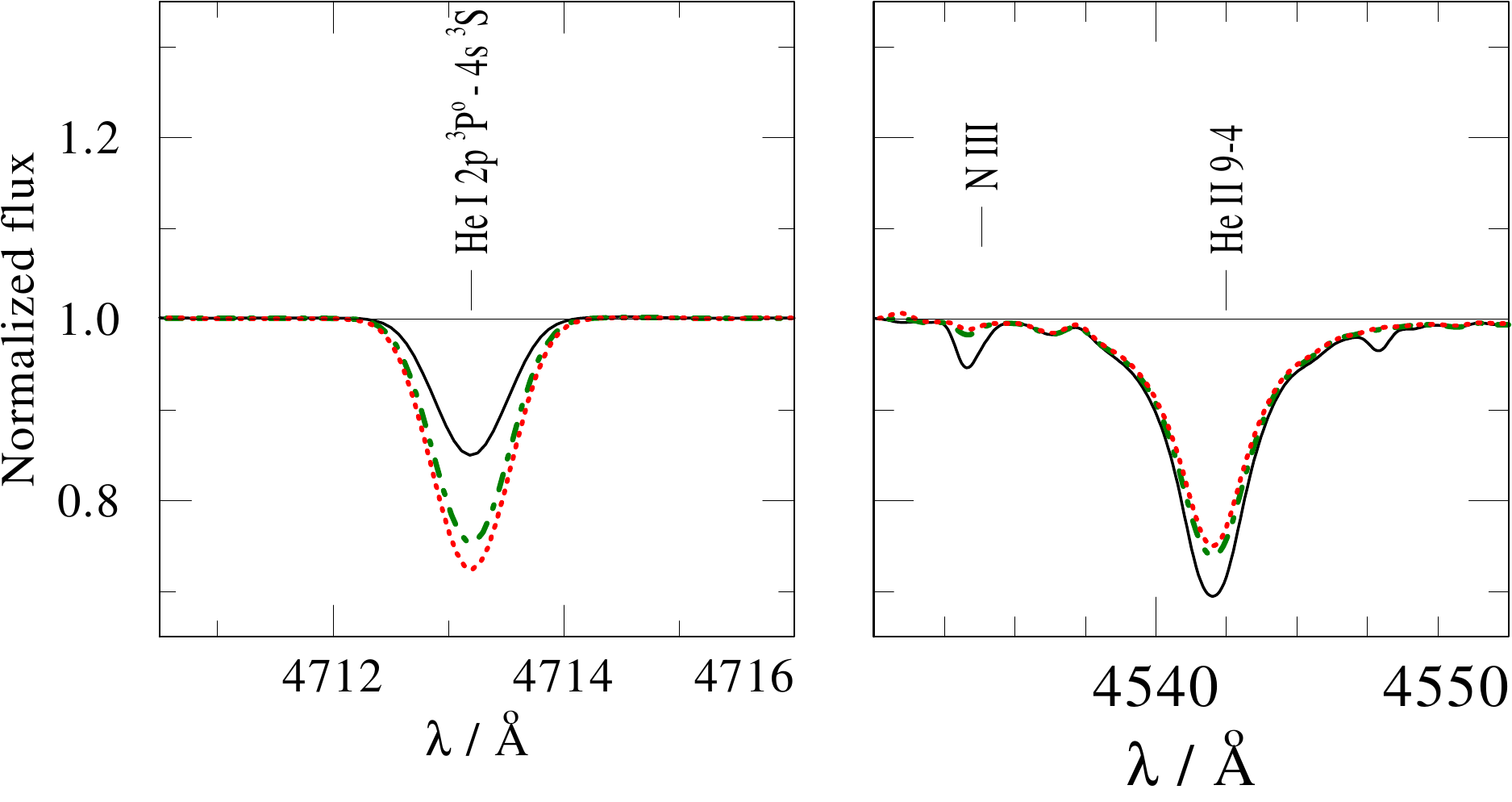}
    \caption{
        As in Fig.\,\ref{fig:spec_comp} but showing zooms on the He\,I\,$\lambda$4713 and He\,II\,$\lambda$4542 line.
    }
    \label{fig:comp_He}
\end{figure}

Figure\,\ref{fig:spec_comp} reveals how the metal lines become weaker with decreasing metallicity. 
A close inspection of this figure, however, also shows that the equivalent widths of the \ion{He}{ii} lines are decreasing with $Z$. A zoom on the \ion{He}{ii}\,$\lambda4542$ line and the \ion{He}{i}\,$\lambda4713$ line is depicted in Fig.\,\ref{fig:comp_He}, revealing that as the \ion{He}{ii} lines get weaker the \ion{He}{i} lines simultaneously become stronger with decreasing $Z$. Although this effect is relatively small for the \ion{He}{ii} lines, it can have a noticeable impact on the parameters that one would deduce from spectral line fits using these models. 
This effect is not limited to the helium lines. Test calculations revealed that it is a general trend that is also displayed by other elements. For example, if the carbon abundance is kept constant but the iron abundance is changed from its default values in the grids to zero, the same effect 
is also visible in the carbon lines. We are confident that this is not a PoWR specific artefact, since the same effect is also shown by the TLUSTY models calculated by \cite{Lanz2003}.

The reason for the observed dependence of the \ion{He}{i} to \ion{He}{ii} line ratios on metallicity is the changing flux level in the UV and extreme UV, which depends on the metal abundances used in the model calculations. 
According to the flux level, the ionization structure of the models shifts to a different balance because of the extreme non-LTE situation within the atmospheres of these stars. This leads to the observed differences in the emergent \element{He} spectra. 
Despite this general mechanism, 
it was not possible to identify specific wavelength ranges or specific transitions that might be chiefly responsible for the observed change in the ionization stratification. Because of the millions of transitions and the various non-LTE effects involved, this is a very difficult task; it is beyond the scope of this paper but deserves a specific study.

Massive stars are found to have an earlier spectral type and appear to be younger at low metallicities compared to their solar companions \citep[e.g.,][]{Massey2004,Martins2004,Mokiem2004,Crowther2006b}. This is because the stars are considered to be more compact at low $Z$. The finding, illustrated in Fig.\,\ref{fig:comp_He}, might appear to contradict this canonical perception. However, the effect presented here is a different one, since the models have the same $R_\ast$ and the difference between the $T_{2/3}$ (effective temperature at $\tau = 2/3$) values of these models is negligible. 
Using  
our models to analyze stars would actually also result in higher temperatures at low $Z$ compared to solar metallicities. 
To understand why, one may imagine two stars, one from the MW and one from the SMC, that have the same spectral type and that exhibit the same equivalent widths in the \ion{He}{i} and \ion{He}{ii} lines. If those lines were to be reproduced by a MW model with a certain stellar temperature, the corresponding model from the SMC grid would not fit to the observations. To reproduce the spectra with a model from the SMC grid, one actually would have to choose a model with a higher $T_\ast$ compared to the MW grid to compensate for the weaker \ion{He}{ii} and stronger \ion{He}{i} lines. 

The observed changes in the \element{He} spectra with the metallicity highlights the need for non-LTE atmosphere models for the spectral analyses of not only OB-type stars but in principle all hot stars. 
This is also evident from Fig.\,\ref{fig:comp_sed}, which compares the SEDs of the models shown in Fig.\,\ref{fig:spec_comp} with a black body of the same effective temperature. While the flux of the models in the IR and beyond is approximated quite well by the black body, the deviations in the UV and extreme UV are huge. 
The black-body SED overestimates the number of hydrogen ionizing photons ($\lambda < 912$\,\AA) by almost 50\% in the selected examples for all three metallicities. The number of \ion{He}{i} ionizing photons ($\lambda < 504$\,\AA) is very low in our detailed models, because these photons are mainly absorbed within the atmosphere and cannot emerge. The black body therefore  over-estimates their number by orders of magnitude. The models selected as examples in Figs.\,\ref{fig:spec_comp}-\ref{fig:comp_He} and Table\,\ref{tab:feedback} are not hot enough to emit photons that can ionize \ion{He}{ii} ($\lambda < 228$\,\AA).
However, for the hottest models in our grids such photons are predicted in significant number. Of course, black bodies completely fail to approximate this part of the spectrum. 
All these examples show that the SEDs of massive stars cannot be approximated with black bodies. Instead, sophisticated stellar atmosphere models are required for the investigation of the radiative feedback of massive stars (see e.g., \citealt{Unsoeld1968} and \citealt{Mihalas1978} for details on stellar atmospheres and the physical background).

\begin{figure}[tbp]
    \centering
    \includegraphics[width=\hsize]{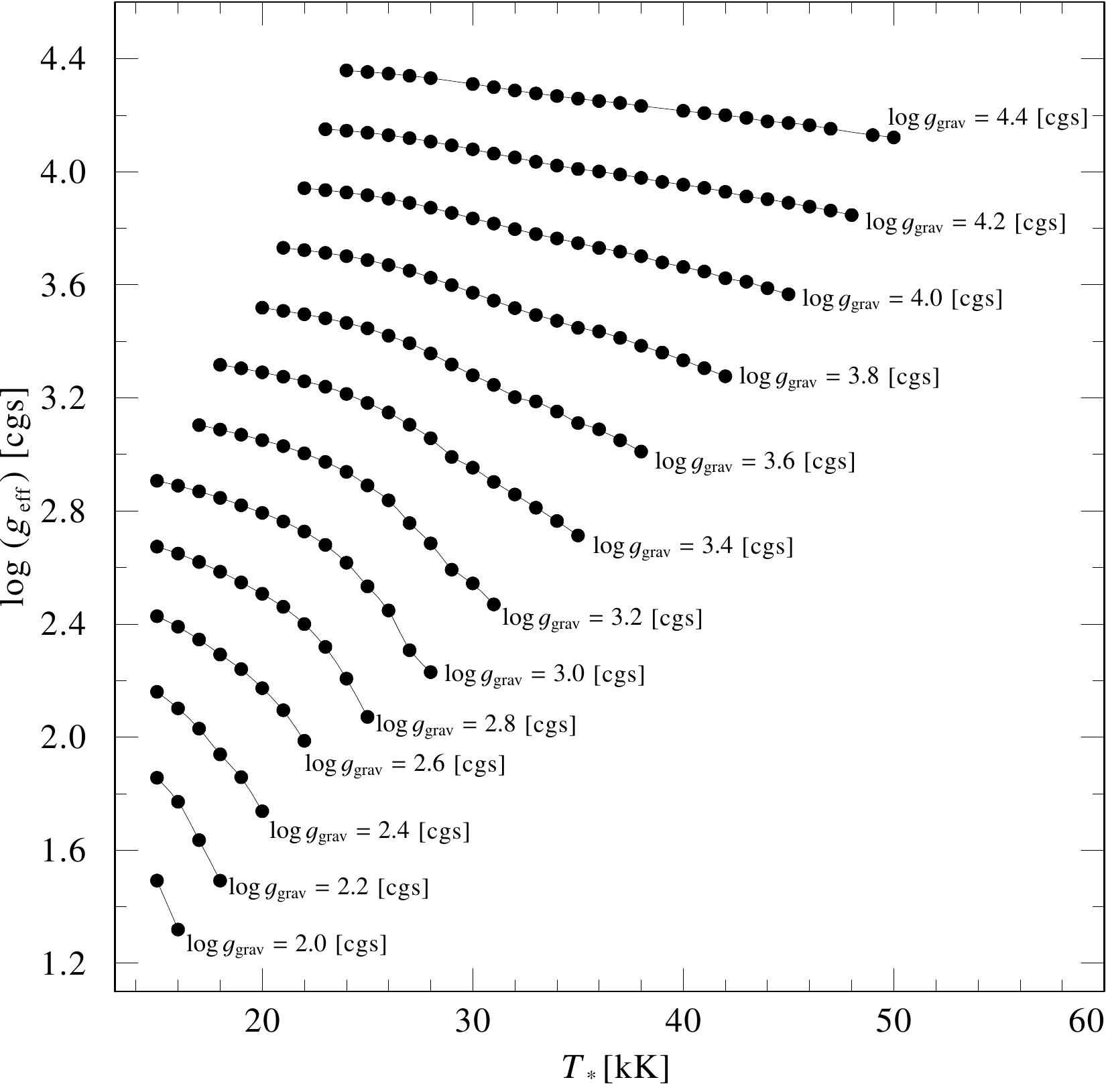}
    \caption{
        $\log g_\mathrm{eff} - T_\ast$ plane of the LMC model grid illustrating the effect of the radiation pressure on the effective surface gravity. Each black dot refers to one grid model. The thin lines connect models with the same $\log g_\mathrm{grav}$. 
    }
    \label{fig:t-g_comp}
\end{figure}

\begin{figure}[tbp]
    \centering
    \includegraphics[width=\hsize]{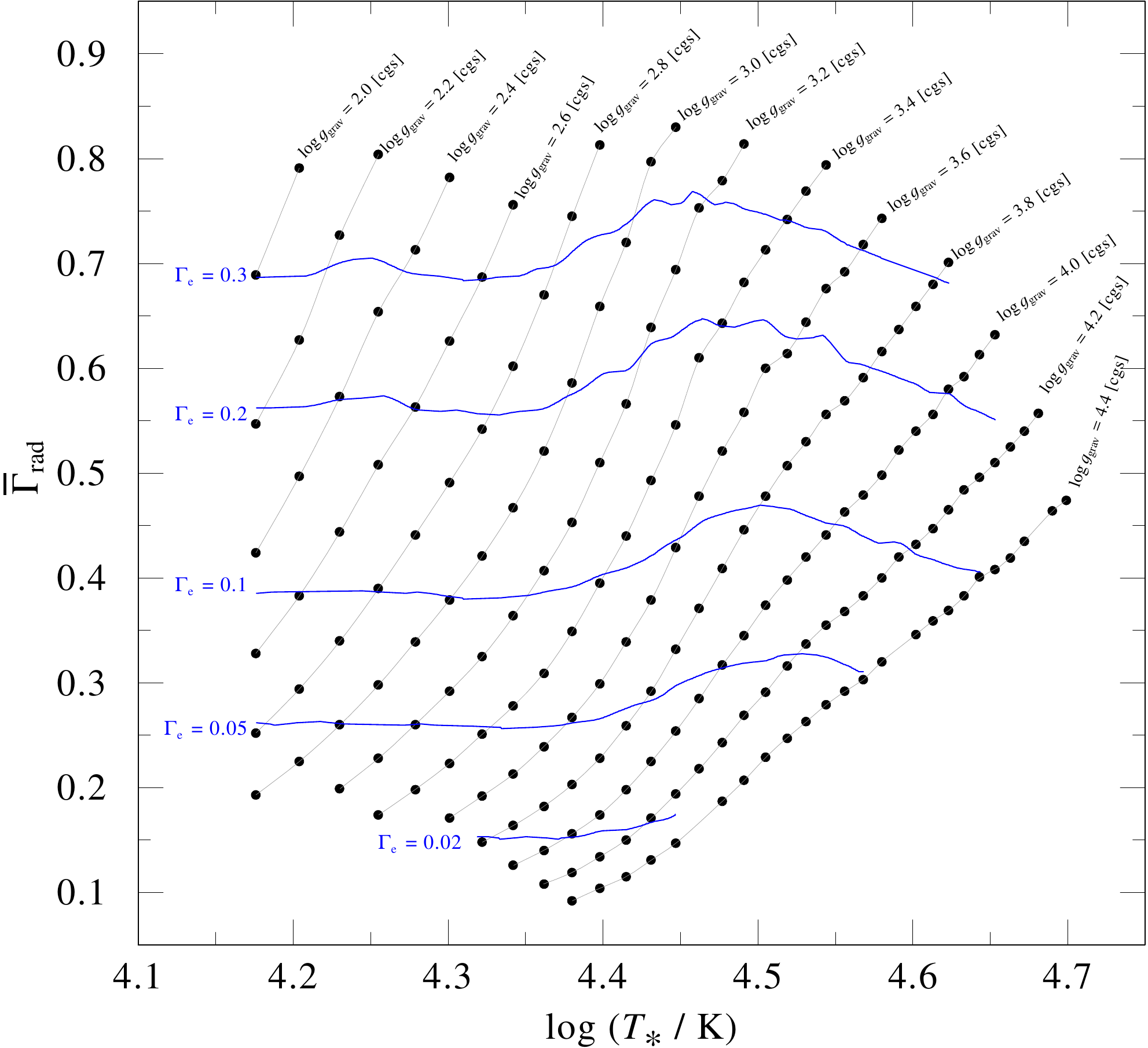}
    \caption{
        Eddington Gamma $\overline{\Gamma}_\mathrm{rad}$ (black dots) plotted vs. stellar temperature on a logarithmic scale for the models from the LMC grid. The thin black lines connect models with the same $\log g_\mathrm{grav}$, while the blue contours depict lines of constant classical Eddington Gamma $\Gamma_{e}$, as labeled. 
    }
    \label{fig:gammarad}
\end{figure}

The mass of a star, $M_\mathrm{spec}$, can be derived spectroscopically 
by fitting the synthetic spectrum to the wings of pressure-broadened
lines. In the case of OB stars, the Balmer lines are specially suitable
for this purpose. The shape and strength of these line wings depend on
the electron pressure at their formation depth, which in OB star
atmospheres is located in the lower, quasi-hydrostatic part of the
atmosphere. However, it is not only gravity that enters the hydrostatic
equation. In fact, the atmospheric pressure is determined by the {\em
effective gravity} $g_\mathrm{eff}$, which is the gravitational
acceleration reduced by the effect of the outward-directed radiation
pressure. 

Hence, the quantity which is measured from fitting the
line wings is $g_\mathrm{eff}$ (see Eq.\,(\ref{eq:geff})), and only with
the proper correction for the radiation pressure  can the correct
spectroscopic mass be obtained. The relation between
$g_\mathrm{eff}$ and $g_\mathrm{grav}$ can be investigated from our
model grids. 
In Fig.\,\ref{fig:t-g_comp} we plot the effective
surface gravity versus the stellar temperature of the models from the
LMC grid.  The effective surface gravity accounts for the full
radiation pressure and is given by
\begin{equation}
\label{eq:geff}
g_\mathrm{eff} = g_\mathrm{grav} (1 - \overline{\Gamma}_\mathrm{rad}),
\end{equation} 
where $g_\mathrm{grav}$ is given by Eq.\,(\ref{eq:ggrav}), and $\overline{\Gamma}_\mathrm{rad}$ is a weighted mean of the full Eddigton Gamma $\Gamma_\mathrm{rad}$ over the hydrostatic domain of the stellar atmosphere as defined by Eq.\,27 in \citet{Sander2015}. In Fig.\,\ref{fig:t-g_comp}, models with the same $\log g_\mathrm{grav}$ are connected by a thin black line.  This figure depicts the difference between $g_\mathrm{grav}$ and $g_\mathrm{eff}$ throughout the grid.
The higher the $L/M$ ratio, the stronger  this effect becomes. This is already evident from the definition of the classical Eddington Gamma
\begin{equation}
\label{eq:Gammae}
\Gamma_\mathrm{e} = \frac{\sigma_\mathrm{e}}{4\pi c G} q_\mathrm{ion}
\frac{L}{M_\ast},
\end{equation}
where $q_\mathrm{ion}$ is the ionization parameter and
$\sigma_\mathrm{e}$ denotes the Thomson opacity. Since $q_\mathrm{ion}$
is not vastly varying throughout the grid, the variation in
$\Gamma_\mathrm{e}$ is mainly due to different $L/M$ ratios.

\begin{figure}[tbp]
    \centering
    \includegraphics[width=\hsize]{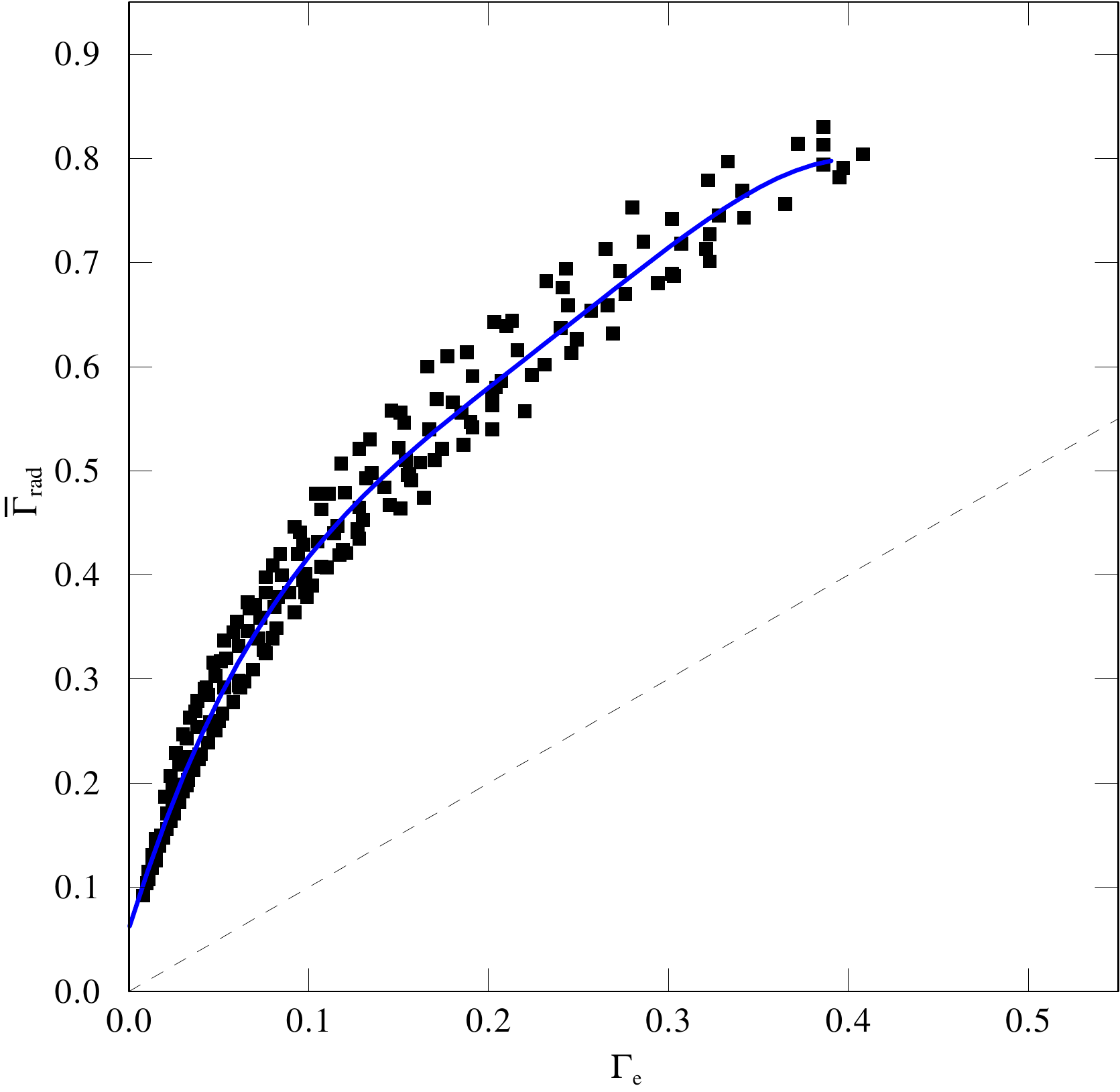}
    \caption{
        Eddington Gamma $\overline{\Gamma}_\mathrm{rad}$ as a function of the classical Eddington Gamma $\Gamma_\mathrm{e}$. Each symbol refers to one model from the LMC grid. The influence of the model parameter $T_\ast$ and $g_\mathrm{eff}$ is limited, as reflected by the modest scatter of the data points. The blue straight line represents the fit to the data points (see Eq.\,(\ref{eq:gamma_fit}) and Table\,\ref{tab:gamma_fit}).  
        The black dashed line indicates $\overline{\Gamma}_\mathrm{rad} = \Gamma_\mathrm{e}$, i.e., the radiation pressure would be purely due to electron scattering. 
    }
    \label{fig:comp_gamma}
\end{figure}

The classical Eddington Gamma $\Gamma_\mathrm{e}$ accounts only for the radiative acceleration due to Thomson scattering by free electrons. 
The full Eddington Gamma ${\Gamma}_\mathrm{rad}$, accounting for all continuum and line opacities, that is, ${\Gamma}_\mathrm{rad} = \Gamma_\mathrm{e} + \Gamma_\mathrm{lines} + \Gamma_\mathrm{true\,\,cont}$, is significantly larger than $\Gamma_\mathrm{e}$. This is illustrated in Figs.\,\ref{fig:gammarad} and \ref{fig:comp_gamma}.   
Figure\,\ref{fig:gammarad} illustrates the connection between the stellar temperature, the full mean Eddington Gamma $\overline{\Gamma}_\mathrm{rad}$, and the classical Eddington Gamma $\Gamma_\mathrm{e}$.
As in Fig.\,\ref{fig:t-g_comp}, the models are taken from the LMC grid. Each filled circle refers to one model, while those models with the same $\log g_\mathrm{grav}$ are connected by a thin black line. The blue contours in this plot refer to lines of the same $\Gamma_\mathrm{e}$. Comparing these contour lines with the $\overline{\Gamma}_\mathrm{rad}$ values demonstrates the idea that a low value for the classical Eddington Gamma $\Gamma_\mathrm{e}$ does not necessarily mean that a star is far from the Eddington limit.
This comparison also indicates that the relation between $\overline{\Gamma}_\mathrm{rad}$ and $\Gamma_\mathrm{e}$ is not linear but quite complex throughout the grid, which is because of the $\overline{\Gamma}_\mathrm{rad}$ temperature dependence. 
This result suggests that stellar properties (e.g., $\dot{M}$) should be correlated with $\Gamma_\mathrm{rad}$ rather than with $\Gamma_\mathrm{e}$. 

We therefore derive the dependence of $\overline{\Gamma}_\mathrm{rad}$ on $\Gamma_\mathrm{e}$. For this purpose, we plot in Fig.\,\ref{fig:comp_gamma} the values of $\overline{\Gamma}_\mathrm{rad}$ over $\Gamma_\mathrm{e}$ for the models from the LMC model grid.
The relation between $\Gamma_\mathrm{e}$ and $\overline{\Gamma}_\mathrm{rad}$ can be best approximated with a fourth-order polynomial of the form
\begin{equation}
\label{eq:gamma_fit}
\overline{\Gamma}_\mathrm{rad} = C_1 + C_2 \Gamma_\mathrm{e} + C_3 \Gamma_\mathrm{e}^2 + C_4 \Gamma_\mathrm{e}^3 + C_5 \Gamma_\mathrm{e}^4\,.
\end{equation}
The coefficients for the fit are given in Table\,\ref{tab:gamma_fit}, where we also include the relations derived by means of the models from the SMC and MW grid. The corresponding figures showing the MW and the SMC fit are shown in Appendix\,\ref{sec:addfigures}. 
In comparison to the LMC relation, the fits to the SMC and MW models lie slightly below and above, respectively, revealing that $Z$ has only a modest effect on  $\overline{\Gamma}_\mathrm{rad}$ within the parameter range studied in this work. While this might sound surprising initially, one must keep in mind that $\overline{\Gamma}_\mathrm{rad}$ is only the mean over the hydrostatic domain \citep[see][]{Sander2015} and does not cover the wind where the influence of $Z$ might be much larger.  

\begin{table}
    \caption{Coefficients of relations between $\Gamma_\mathrm{e}$ and $\overline{\Gamma}_\mathrm{rad}$ (Eq.\,\ref{eq:gamma_fit}) for the SMC, LMC, and MW models.}
    \label{tab:gamma_fit}      
    \centering
    \begin{tabular}{lSSSSS}
        \hline\hline 
        \rule[0mm]{0mm}{2.2ex} 
        Grid & $C_{1}$  & $C_{2}$ & $C_{3}$ & $C_{4}$ & $C_{5}$ \\
        \hline 
        SMC & 0.06 & 4.69 & -19.93 & 51.97 & -51.13 \rule[0mm]{0mm}{2.5ex} \\
        LMC & 0.06 & 5.57 & -26.68 & 74.77 & -78.33 \\
        MW  & 0.08 & 5.26 & -19.88 & 40.06 & -29.65 \\
        \hline 
    \end{tabular}
\end{table}

\begin{figure}[tbp]
    \centering
    \includegraphics[angle=-90,width=\hsize]{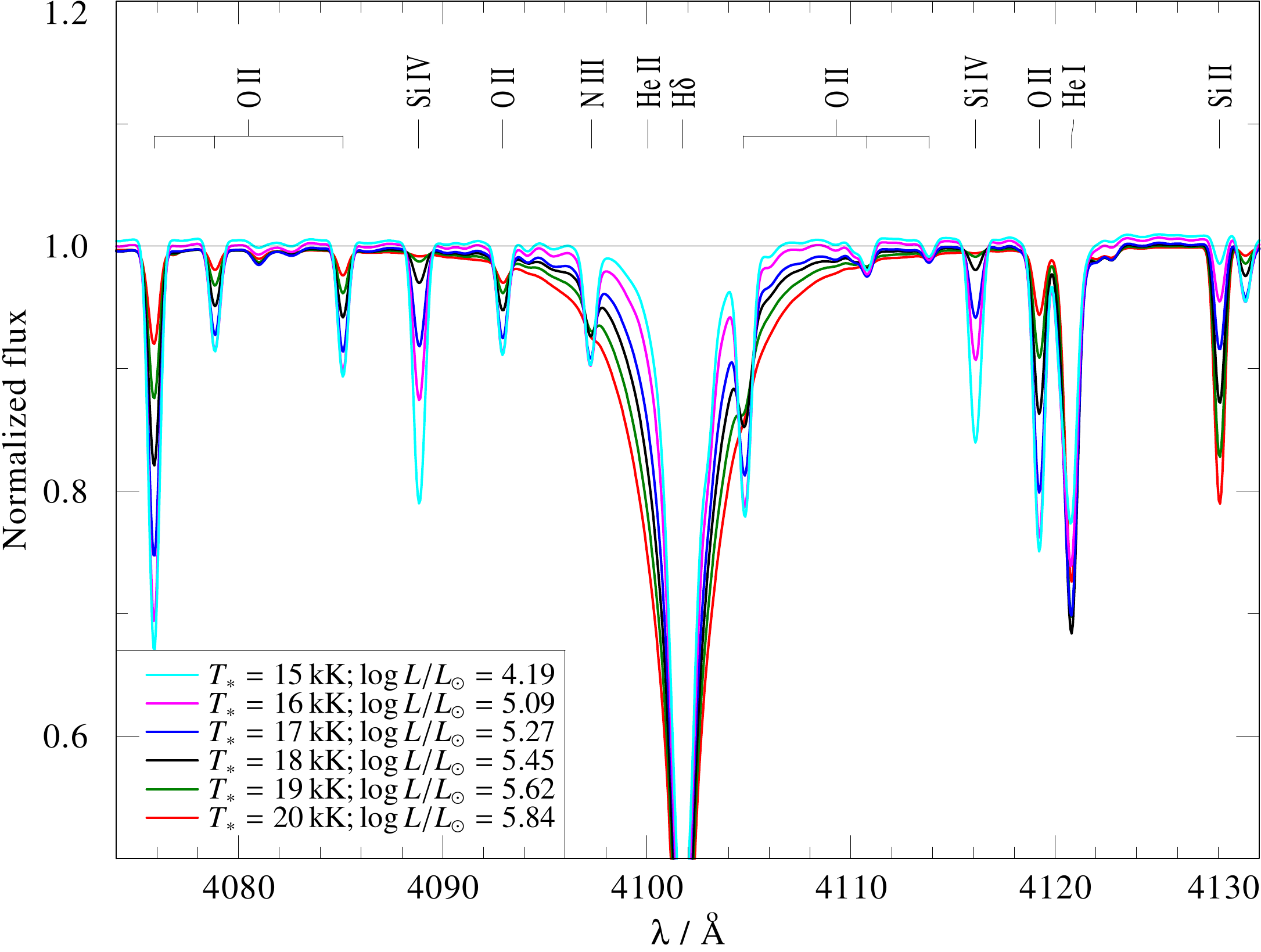}
    \caption{
        Normalized spectra of six models from the LMC grid showing the spectral range around  the H\,$\delta$ line. All models have the same $\log g_\mathrm{grav}$ but different $T_\ast$ and $L$. See inlet for details. 
    }
    \label{fig:comp_Hdelta}
\end{figure}

The effect of the radiation pressure on the Balmer line wings is illustrated 
in Fig.\,\ref{fig:comp_Hdelta}, which shows the spectral region around
the H\,$\delta$ line for six different models from the LMC grid. These
models have the same value for the pure gravitational acceleration of
$\log g_\mathrm{grav} = 2.4$, but exhibit substantial differences in
the pressure broadened H\,$\delta$ line wings. This is because of the
different $\log g_\mathrm{eff}$ values that vary between 1.7\,[cgs] and
2.2\,[cgs] due to the change in the radiation pressure. The variations
in the other lines visible in Fig.\,\ref{fig:comp_Hdelta} are mainly
attributable to the different stellar temperatures of the models.  

\begin{figure}[tbp]
    \centering
    \includegraphics[angle=-90,width=\hsize]{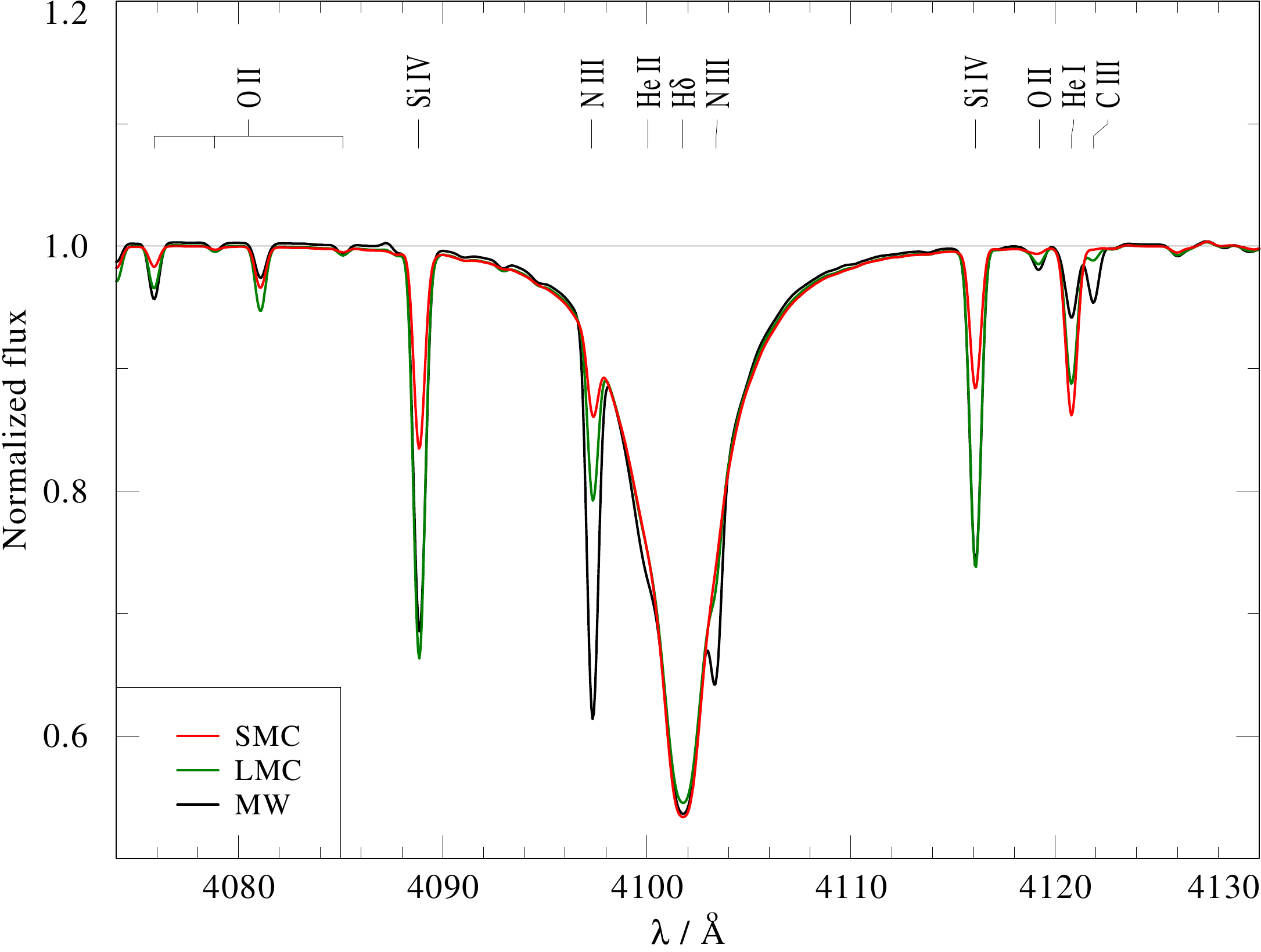}
    \caption{
        Like Fig.\,\ref{fig:comp_Hdelta} but for H$\delta$, showing the models underlying Figs.\,\ref{fig:spec_comp} and \ref{fig:comp_He}.
    }
    \label{fig:comp_Hdelta_2}
\end{figure}

As shown above for $\overline{\Gamma}_\mathrm{rad}$, the impact of the metallicity on the density structure and the pressure broadening of the spectral lines is quite weak in the metallicity domain explored in this work. 
This is illustrated by Fig.\,\ref{fig:comp_Hdelta_2} that displays the same spectral range as Fig.\,\ref{fig:comp_Hdelta}, while it depicts the models shown in Fig.\,\ref{fig:spec_comp}. These models exhibit the same stellar parameters, but were calculated for MW, LMC, and SMC metallicity. The small difference between the wings of the H\,$\delta$ line exemplifies the limited effect of the metallicity. 

\section{Potential applications}
\label{sect:summary}

We have presented extensive atmosphere model grids for OB-type stars calculated with the PoWR code for MW, LMC, and SMC metallicities. Altogether 785 models have been calculated for four model grids. Two grids are available for SMC metallicities, while one grid has so far been calculated for the MW and another for the LMC. Further grids extending the parameter space, especially with respect to the mass-loss rate, are in preparation and will be the subject of a forthcoming paper discussing calibrations between spectral types and physical parameters.

Based on these models, we have illustrated the impact of the radiation pressure on the surface gravity and on the emergent spectra. We derived approximate relations between the classical Eddington Gamma, accounting only for scattering by free electrons, and the full Eddington Gamma, which takes all continuum and line opacities into account. 

The immediate application of the model grids provided here is for
quantitative spectral analyses. Such analysis proceeds in two steps.   
First, the observed (normalized or flux-calibrated) line spectrum is
fitted to the synthetic spectra from the grid. The stellar temperature
can be deduced by fitting the helium and metal lines, 
paying special attention to the temperature-sensitive ratios between 
lines  of different ionization stages. The surface gravity is adjusted
by fitting the pressure-broadened profiles, especially of the hydrogen 
and helium lines. The turbulent and rotational contribution to the line
broadening must be separated, for example\ with the \texttt{iacob-broad} tool
\citep{Simon-Diaz2014} applied to narrow metal lines. 

The UV resonance lines, and possibly the strongest lines in the optical 
(e.g.,\ H$\alpha$) might form in the stellar wind; comparison with the
grids calculated for different mass-loss rates may thus give a
constraint to this parameter. 

As the second step,  the luminosity of the star is determined
from fitting the model SED  to flux-calibrated spectra and/or filter
photometry. Here, the model flux has to be scaled according to the
distance of the star, that is,\ knowledge of the distance is essential here.
At the same time, the interstellar reddening  and extinction need to be
accounted for, for example by modifying the model SED by means of a reddening
law, so that the shape of the observed SED is reproduced. Thus, this
procedure allows to simultaneously derive the luminosity of a star and
the interstellar reddening along the line of sight.  

Besides spectra and SEDs, further model predictions such as feedback parameters and atmospheric stratifications are provided online for all models as well. These model grids allow a wide range of applications, from spectral analyses to theoretical studies that require atmospheric stratifications of atomic population numbers as input.

\begin{acknowledgements}

We thank the anonymous referee for their constructive comments.
A.\,A.\,C.\,S. is supported by the Deutsche Forschungsgemeinschaft (DFG) under grant HA 1455/26. 
V.\,R. is grateful for financial support from the Deutsche Akademische Austauschdienst (DAAD) as part of the Graduate School Scholarship Program. 
T.\,S. and L.\,M.\,O. acknowledge support from the german "Verbundforschung" (DLR) grants, 50\,OR\,1612 and 50\,OR\,1508, respectively. 

\end{acknowledgements}

\bibliographystyle{aa}
\bibliography{paper}



\begin{appendix} 
\section{Additional tables}
\label{sec:addtables}

\begin{table*}[htbp]
\caption{Atomic model used to construct the OB-type model grids} 
\label{table:model_atoms}
\centering  
\begin{tabular}{lSS|lSS}
\hline\hline 
   Ion &
   \multicolumn{1}{c}{Number of levels} &
   \multicolumn{1}{c}{Number of lines\tablefootmark{a}}  &
   Ion & 
   \multicolumn{1}{c}{Number of levels} &
   \multicolumn{1}{c}{Number of lines\tablefootmark{a}}  \rule[0mm]{0mm}{3.5mm} \\
\hline  
   \ion{H}{i}    & 22 & 231 & \ion{S}{v}                     & 10 &   8 \rule[0mm]{0mm}{4.0mm}  \\
   \ion{H}{ii}   & 1  &   0 & \ion{S}{vi}                    &  1 &   0 \\
   \ion{He}{i}   & 35 & 271 & \ion{Mg}{i}                    &  1 &   0 \\
   \ion{He}{ii}  & 26 & 325 & \ion{Mg}{ii}                   & 32 & 120 \\
   \ion{He}{iii} &  1 &   0 & \ion{Mg}{iii}                  & 43 & 158 \\
   \ion{N}{i}    & 10 &  13 & \ion{Mg}{iv}                   & 17 &  27 \\
   \ion{N}{ii}   & 38 & 201 & \ion{Mg}{v}                    & 20 &  25 \\
   \ion{N}{iii}\tablefootmark{b} & 56\,85 & 219 \,464 & \ion{Si}{i}\tablefootmark{b} & 20 & 43 \,45 \\
   \ion{N}{iv}   & 38 & 154 & \ion{Si}{ii}                   & 20 &  35 \\
   \ion{N}{v}    & 20 & 114 & \ion{Si}{iii}                  & 24 &  68 \\
   \ion{N}{vi}   & 14 &  48 & \ion{Si}{iv}                   & 23 &  72 \\
   \ion{C}{i}    & 15 &  30 & \ion{Si}{v}                    &  1 &   0 \\
   \ion{C}{ii}   & 32 & 148 & \ion{P}{iv}                    & 12 &  16 \\
   \ion{C}{iii}  & 40 & 226 & \ion{P}{v}                     & 11 &  22 \\
   \ion{C}{iv}   & 25 & 230 & \ion{P}{vi}                    &  1 &   0 \\
   \ion{C}{v}    & 29 & 120 & \ion{G}{i}\tablefootmark{c}    &  1 &   0 \\
   \ion{C}{vi}   &  1 &   0 & \ion{G}{ii}\tablefootmark{c}   &  3 &   2 \\
   \ion{O}{i}    & 13 &  15 & \ion{G}{iii}\tablefootmark{c}  & 13 &  40 \\
   \ion{O}{ii}   & 37 & 150 & \ion{G}{iv}\tablefootmark{c}   & 18 &  77 \\
   \ion{O}{iii}  & 33 & 121 & \ion{G}{v}\tablefootmark{c}    & 22 & 107 \\
   \ion{O}{iv}   & 29 &  76 & \ion{G}{vi}\tablefootmark{c}   & 29 & 194 \\
   \ion{O}{v}    & 36 & 153 & \ion{G}{vi}\tablefootmark{c}   & 29 & 194 \\
   \ion{O}{vi}   & 16 & 101 & \ion{G}{vii}\tablefootmark{c}  & 19 &  87 \\
   \ion{O}{vii}  & 15 &  64 & \ion{G}{viii}\tablefootmark{c} & 14 &  49 \\ 
   \ion{S}{iii}  & 23 &  38 & \ion{G}{ix}\tablefootmark{c}   & 15 &  56 \\
   \ion{S}{iv}   & 11 &  13 & \\
\hline 
\end{tabular}
\tablefoot{
\tablefoottext{a}{Number of lines accounted for during the comoving-frame calculations.}
\tablefoottext{b}{For those quantities where two values are given, the second one refers to the Galactic OB-star grid, while the first one is valid for all other grids.}
\tablefoottext{c}{G denotes a generic atom which incorporates the following iron group elements: \element{Sc}, \element{Ti}, \element{V}, \element{Cr}, \element{Mn}, \element{Fe}, \element{Co}, and \element{Ni}. The  corresponding ions are treated by means of a superlevel approach \citep[for details see][]{Graefener2002}.}
}
\end{table*}

\input{lmc-ob-i_25-32_stratification.tex}

\section{Additional figures}
\label{sec:addfigures}

\begin{figure}[tbp]
    \centering
    \includegraphics[width=\hsize]{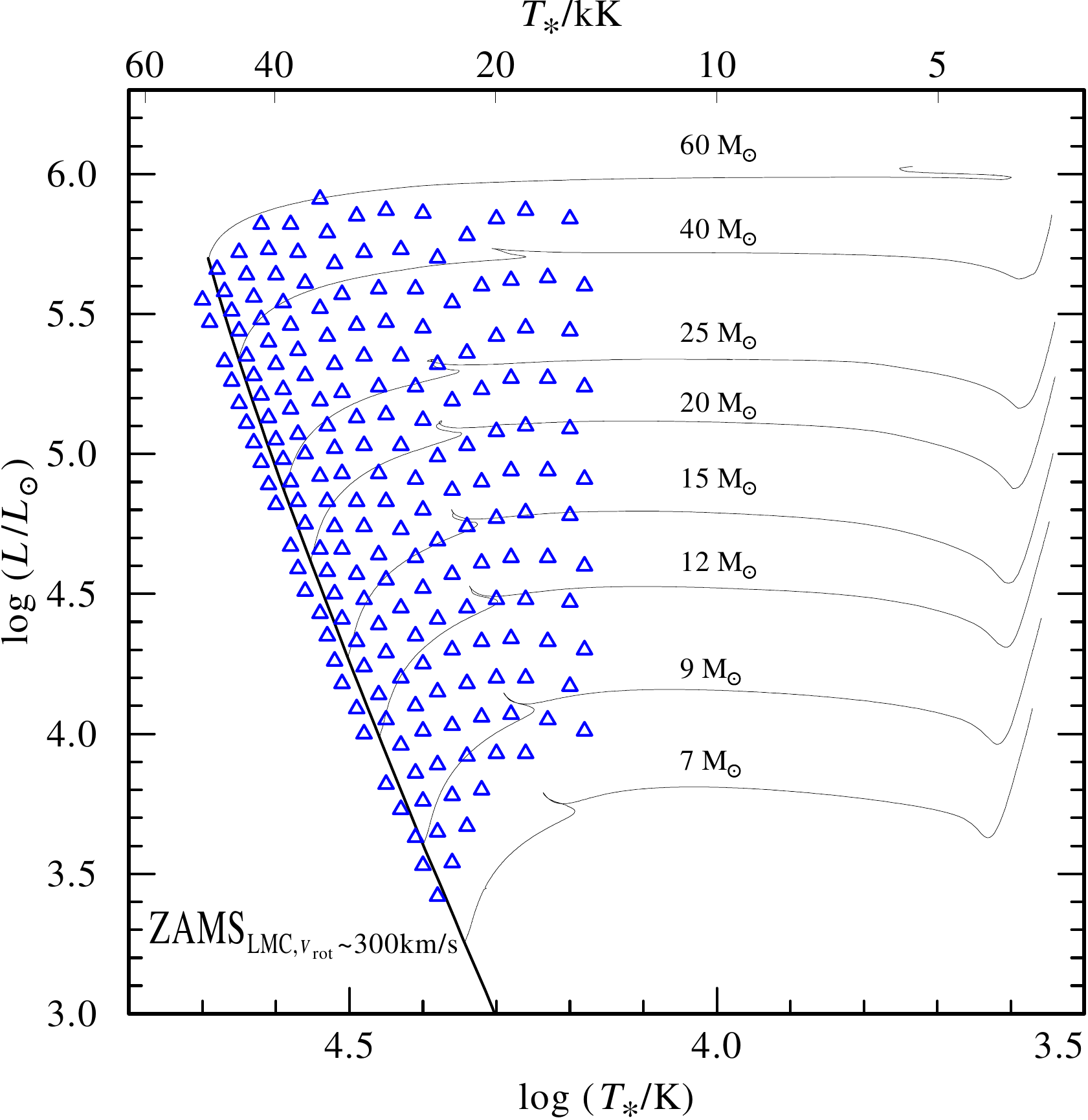}
    \caption{
        Same as Fig.\,\ref{fig:hrd_smc} but for the LMC grid.  
    }
    \label{fig:hrd_lmc}
\end{figure}

\begin{figure}[tbp]
    \centering
    \includegraphics[width=\hsize]{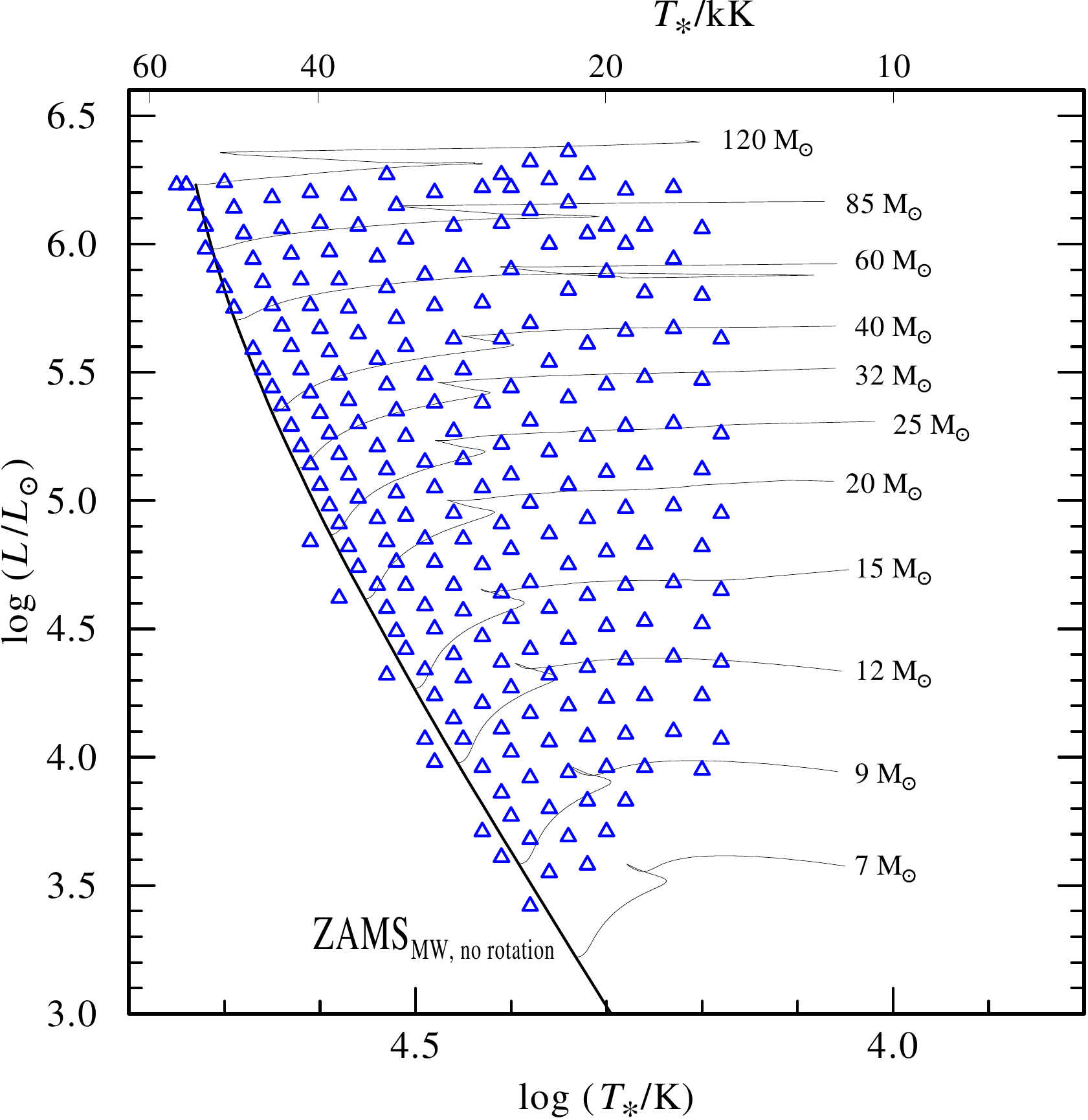}
    \caption{
        Same as Fig.\,\ref{fig:hrd_smc} but for the MW grid. The depicted stellar evolution tracks were calculated by \citet{Ekstroem2012}. Only the relevant parts of the tracks are plotted.  
    }
    \label{fig:hrd_solar}
\end{figure}

\begin{figure}[tbp]
    \centering
    \includegraphics[width=\hsize]{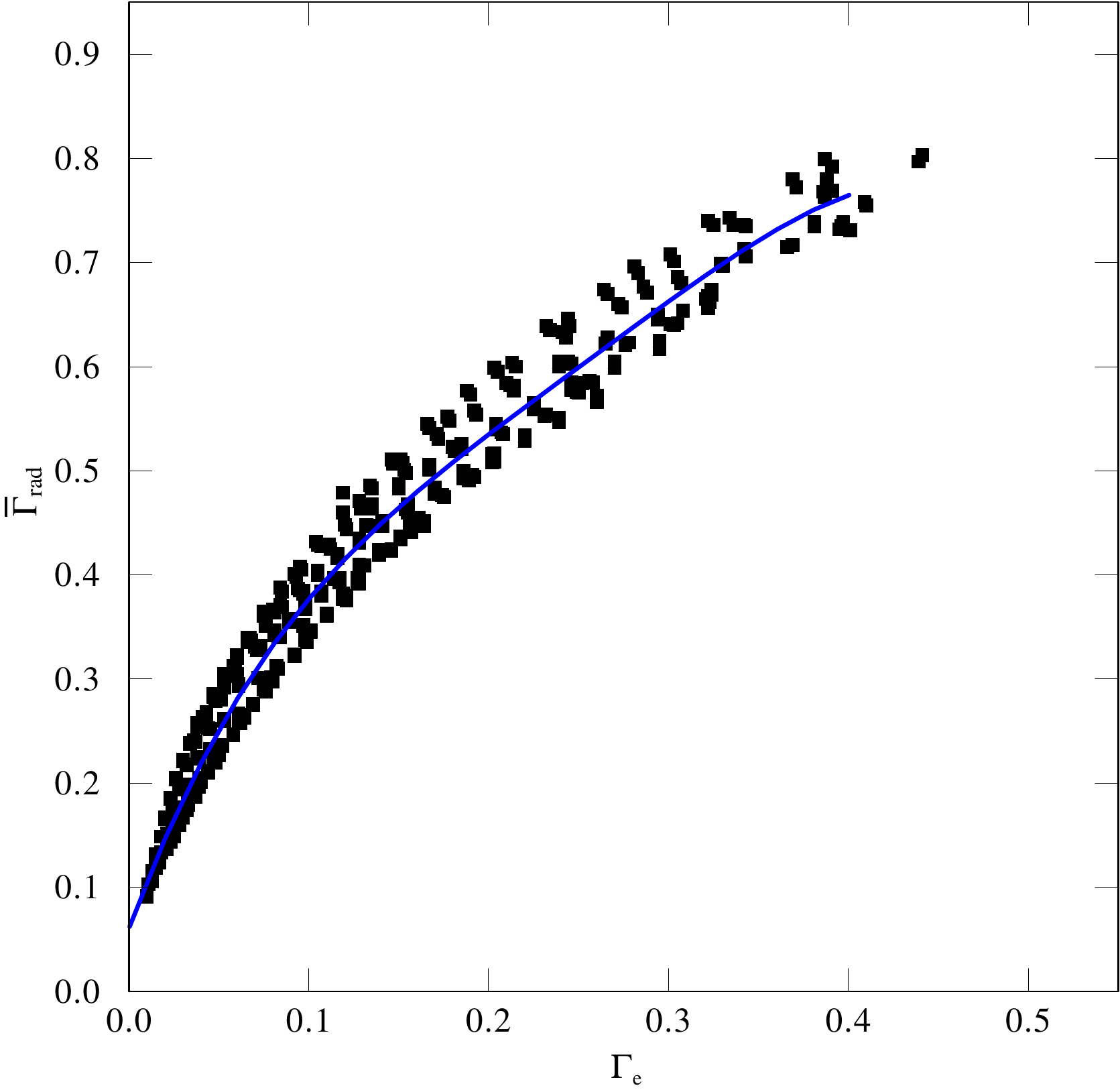}
    \caption{
        Same as Fig.\,\ref{fig:comp_gamma} but for the models from the SMC grid. 
    }
    \label{fig:comp_gamma_smc}
\end{figure}

\begin{figure}[tbp]
    \centering
    \includegraphics[width=\hsize]{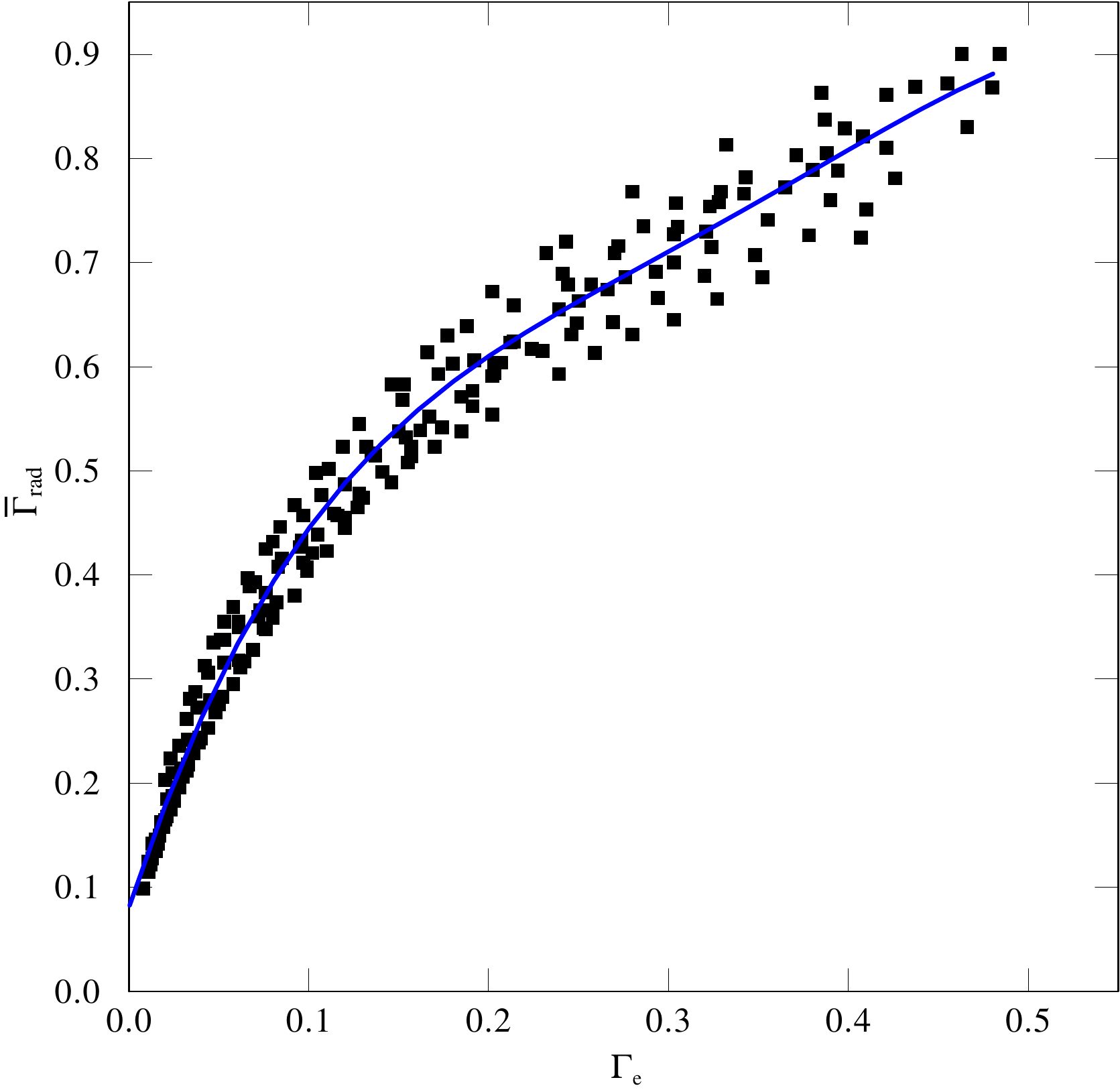}
    \caption{
        Same as Fig.\,\ref{fig:comp_gamma} but for the models from the MW grid.
    } 
    \label{fig:comp_gamma_mw}
\end{figure}

\end{appendix} 

\end{document}

%% file: lmc-ob-i_25-32_stratification.tex
\longtab[2]{
\small
\begin{landscape}
\begin{longtable}{ccScccccccSSc}
    \caption{Atmospheric structure of the model with $T_\ast = 25\,\mathrm{kK}$, $\log g = 3.2\,\mathrm{[cgs]}$, $\log \dot{M}/(M_\odot/\mathrm{yr}) = -7.0$, $\log L/L_\odot = 5.12$, $R_\ast = 19.4\,R_\odot$ and $v_\infty = 1578\,\mathrm{km/s}$ from the LMC model grid} 
    \label{table:structure}\\
        \hline\hline 
        \multicolumn{1}{c}{Depth}  & \multicolumn{1}{c}{$r-1$}     & \multicolumn{1}{c}{$\log (r-1)$}   & \multicolumn{1}{c}{$T_\mathrm{e}$} & \multicolumn{1}{c}{$\log N$}                           & \multicolumn{1}{c}{$\log N_\mathrm{e}$}                & \multicolumn{1}{c}{$\tau_\mathrm{Thom}$} & \multicolumn{1}{c}{$\tau_\mathrm{Ross}$} & \multicolumn{1}{c}{$\tau_\mathrm{Ross, cont}$} & \multicolumn{1}{c}{$v$}        & \multicolumn{1}{c}{$\partial v / \partial r$}   & \multicolumn{1}{c}{$D$} & \multicolumn{1}{c}{$\mu$} \rule[0mm]{0mm}{3.5mm} \\
        \multicolumn{1}{c}{index}  & \multicolumn{1}{c}{[$R_\ast$]} & \multicolumn{1}{c}{([$R_\ast$])}  & \multicolumn{1}{c}{[K]}                             & \multicolumn{1}{c}{$[\mathrm{Atoms} / \mathrm{cm}^3]$} & \multicolumn{1}{c}{$[\mathrm{Electrons} / \mathrm{cm}^3]$} &                                          & 				              &                                                & \multicolumn{1}{c}{[km\,/\,s]} & \multicolumn{1}{c}{[$\frac{\mathrm{km}\,/\,\mathrm{s}}{R_\ast}$]} & & \multicolumn{1}{c}{[u]}     \\
        \hline 
        \endfirsthead
        \caption{continued.}\\
        \hline
        \hline        
        \multicolumn{1}{c}{Depth}  & \multicolumn{1}{c}{$r-1$}     & \multicolumn{1}{c}{$\log (r-1)$}   & \multicolumn{1}{c}{$T_\mathrm{e}$} & \multicolumn{1}{c}{$\log N$}                           & \multicolumn{1}{c}{$\log N_\mathrm{e}$}                & \multicolumn{1}{c}{$\tau_\mathrm{Thom}$} & \multicolumn{1}{c}{$\tau_\mathrm{Ross}$} & \multicolumn{1}{c}{$\tau_\mathrm{Ross, cont}$} & \multicolumn{1}{c}{$v$}        & \multicolumn{1}{c}{$\partial v / \partial r$}   & \multicolumn{1}{c}{$D$} & \multicolumn{1}{c}{$\mu$} \rule[0mm]{0mm}{3.5mm} \\
        \multicolumn{1}{c}{index}  & \multicolumn{1}{c}{[$R_\ast$]} & \multicolumn{1}{c}{([$R_\ast$])}  & \multicolumn{1}{c}{[K]}                             & \multicolumn{1}{c}{$[\mathrm{Atoms} / \mathrm{cm}^3]$} & \multicolumn{1}{c}{$[\mathrm{Electrons} / \mathrm{cm}^3]$} &                                          & 				              &                                                & \multicolumn{1}{c}{[km\,/\,s]} & \multicolumn{1}{c}{[$\frac{\mathrm{km}\,/\,\mathrm{s}}{R_\ast}$]} & & \multicolumn{1}{c}{[u]}     \\
        \hline
        \endhead
        \hline
        \endfoot
        \hline
        \endlastfoot 
         \rule[0mm]{0mm}{3.5mm}1  &   99.0     &   2.00       &    11092.    &   4.930       &         4.930     &     0.000 &    0.0000000000 &  0.0000000000 &  1578.     &     0.1948  &     10.0000 &    0.6232 \\
         2  &   78.3     &   1.89       &    11359.    &   5.133       &         5.133     &     0.000 &    0.0000021310 &  0.0000020570 &  1575.     &     0.1948  &     10.0000 &    0.6232 \\
         3  &   64.5     &   1.81       &    11635.    &   5.299       &         5.299     &     0.000 &    0.0000042773 &  0.0000041251 &  1571.     &     0.2978  &     10.0000 &    0.6232 \\
         4  &   53.1     &   1.73       &    11917.    &   5.466       &         5.467     &     0.000 &    0.0000069040 &  0.0000066516 &  1567.     &     0.4577  &     10.0000 &    0.6232 \\
         5  &   41.6     &   1.62       &    12250.    &   5.675       &         5.675     &     0.000 &    0.0000110269 &  0.0000106092 &  1561.     &     0.7183  &     10.0000 &    0.6232 \\
         6  &   32.0     &   1.50       &    12572.    &   5.902       &         5.902     &     0.000 &    0.0000168062 &  0.0000161453 &  1552.     &     1.1529  &     10.0000 &    0.6232 \\
         7  &   25.1     &   1.40       &    12823.    &   6.107       &         6.107     &     0.000 &    0.0000234960 &  0.0000225419 &  1542.     &     1.9177  &     10.0000 &    0.6232 \\
         8  &   19.4     &   1.29       &    13046.    &   6.327       &         6.327     &     0.000 &    0.0000327130 &  0.0000313413 &  1528.     &     3.1859  &     10.0000 &    0.6232 \\
         9  &   14.5     &   1.16       &    13255.    &   6.568       &         6.568     &     0.000 &    0.0000460184 &  0.0000440242 &  1508.     &     5.2625  &     10.0000 &    0.6232 \\
        10  &   11.1     &   1.05       &    13435.    &   6.791       &         6.792     &     0.000 &    0.0000620148 &  0.0000592503 &  1484.     &     9.0205  &     10.0000 &    0.6232 \\
        11  &   8.31     &  0.920       &    13613.    &   7.028       &         7.028     &     0.000 &    0.0000842619 &  0.0000803982 &  1452.     &    15.1037  &      9.8695 &    0.6232 \\
        12  &   6.19     &  0.791       &    13793.    &   7.266       &         7.266     &     0.000 &    0.0001135612 &  0.0001082715 &  1410.     &    24.0078  &      7.9828 &    0.6232 \\
        13  &   4.92     &  0.692       &    13968.    &   7.446       &         7.447     &     0.000 &    0.0001412746 &  0.0001346956 &  1371.     &    36.7295  &      6.1332 &    0.6232 \\
        14  &   4.06     &  0.609       &    14084.    &   7.595       &         7.595     &     0.000 &    0.0001685249 &  0.0001607392 &  1332.     &    52.5968  &      4.7827 &    0.6232 \\
        15  &   3.33     &  0.523       &    14144.    &   7.745       &         7.745     &     0.000 &    0.0002010498 &  0.0001918956 &  1287.     &    72.5317  &      3.6831 &    0.6232 \\
        16  &   2.72     &  0.434       &    14192.    &   7.897       &         7.897     &     0.000 &    0.0002398968 &  0.0002291852 &  1235.     &    99.8185  &      2.8425 &    0.6232 \\
        17  &   2.20     &  0.342       &    14296.    &   8.049       &         8.049     &     0.000 &    0.0002862973 &  0.0002737889 &  1174.     &   136.9388  &      2.2287 &    0.6232 \\
        18  &   1.76     &  0.246       &    14550.    &   8.204       &         8.204     &     0.000 &    0.0003417949 &  0.0003271622 &  1104.     &   183.1181  &      1.7959 &    0.6232 \\
        19  &   1.44     &  0.159       &    14927.    &   8.338       &         8.338     &     0.000 &    0.0003980500 &  0.0003812431 &  1035.     &   235.5595  &      1.5343 &    0.6232 \\
        20  &   1.24     &  0.0924      &    15306.    &   8.437       &         8.438     &     0.000 &    0.0004451048 &  0.0004264398 &  980.4     &   290.7219  &      1.3922 &    0.6232 \\
        21  &   1.08     &  0.0353      &    15740.    &   8.521       &         8.521     &     0.001 &    0.0004884130 &  0.0004679811 &  931.6     &   343.0829  &      1.2999 &    0.6232 \\
        22  &  0.958     & -0.0185      &    16205.    &   8.598       &         8.598     &     0.001 &    0.0005316059 &  0.0005093346 &  884.8     &   397.1017  &      1.2325 &    0.6232 \\
        23  &  0.844     & -0.0735      &    16653.    &   8.674       &         8.674     &     0.001 &    0.0005782318 &  0.0005538785 &  836.2     &   458.4503  &      1.1789 &    0.6232 \\
        24  &  0.730     & -0.137       &    17041.    &   8.760       &         8.760     &     0.001 &    0.0006346196 &  0.0006076280 &  780.2     &   527.8112  &      1.1321 &    0.6232 \\
        25  &  0.627     & -0.203       &    17383.    &   8.848       &         8.848     &     0.001 &    0.0006973609 &  0.0006673030 &  721.3     &   605.8287  &      1.0957 &    0.6231 \\
        26  &  0.542     & -0.266       &    17751.    &   8.928       &         8.928     &     0.001 &    0.0007592909 &  0.0007260757 &  666.4     &   693.1416  &      1.0704 &    0.6231 \\
        27  &  0.467     & -0.330       &    18157.    &   9.009       &         9.009     &     0.001 &    0.0008258627 &  0.0007891034 &  611.0     &   790.3683  &      1.0511 &    0.6231 \\
        28  &  0.392     & -0.406       &    18619.    &   9.102       &         9.102     &     0.001 &    0.0009075338 &  0.0008662189 &  547.9     &   898.0311  &      1.0349 &    0.6231 \\
        29  &  0.325     & -0.488       &    19068.    &   9.199       &         9.199     &     0.001 &    0.0009988386 &  0.0009521709 &  483.5     &  1016.4327  &      1.0229 &    0.6231 \\
        30  &  0.273     & -0.563       &    19383.    &   9.288       &         9.288     &     0.001 &    0.0010868222 &  0.0010347657 &  426.9     &  1146.0976  &      1.0154 &    0.6231 \\
        31  &  0.228     & -0.641       &    19637.    &   9.378       &         9.379     &     0.001 &    0.0011802661 &  0.0011222811 &  372.4     &  1287.8731  &      1.0101 &    0.6231 \\
        32  &  0.184     & -0.736       &    19931.    &   9.489       &         9.489     &     0.001 &    0.0012986344 &  0.0012328550 &  311.3     &  1455.3755  &      1.0059 &    0.6231 \\
        33  &  0.139     & -0.856       &    20302.    &   9.630       &         9.630     &     0.001 &    0.0014559466 &  0.0013792277 &  242.8     &  1661.2123  &      1.0029 &    0.6231 \\
        34  &  0.102     & -0.992       &    20568.    &   9.799       &         9.800     &     0.002 &    0.0016491372 &  0.0015580747 &  175.6     &  1907.3123  &      1.0012 &    0.6231 \\
        35  &  0.756E-01 &  -1.12       &    20543.    &   9.982       &         9.982     &     0.002 &    0.0018545980 &  0.0017474464 &  121.1     &  2203.5416  &      1.0004 &    0.6231 \\
        36  &  0.573E-01 &  -1.24       &    20228.    &  10.191       &        10.191     &     0.002 &    0.0020817416 &  0.0019561177 &  77.35     &  2539.2323  &      1.0001 &    0.6231 \\
        37  &  0.460E-01 &  -1.34       &    19749.    &  10.428       &        10.428     &     0.002 &    0.0023216637 &  0.0021757205 &  45.88     &  2914.4143  &      1.0000 &    0.6232 \\
        38  &  0.405E-01 &  -1.39       &    19290.    &  10.641       &        10.641     &     0.002 &    0.0025172592 &  0.0023538750 &  28.35     &  3342.2776  &      1.0000 &    0.6232 \\
        39  &  0.378E-01 &  -1.42       &    18958.    &  10.831       &        10.831     &     0.003 &    0.0026740556 &  0.0024957345 &  18.42     &  3771.8949  &      1.0000 &    0.6232 \\
        40  &  0.363E-01 &  -1.44       &    18737.    &  11.009       &        11.009     &     0.003 &    0.0028087842 &  0.0026165247 &  12.27     &  4166.1658  &      1.0000 &    0.6232 \\
        41  &  0.351E-01 &  -1.46       &    18523.    &  11.251       &        11.251     &     0.003 &    0.0029899661 &  0.0027763816 &  7.035     &  3618.5663  &      1.0000 &    0.6232 \\
        42  &  0.339E-01 &  -1.47       &    18284.    &  11.511       &        11.511     &     0.003 &    0.0032981372 &  0.0030419152 &  3.875     &  2281.9209  &      1.0000 &    0.6232 \\
        43  &  0.330E-01 &  -1.48       &    18066.    &  11.724       &        11.724     &     0.003 &    0.0037550243 &  0.0034258625 &  2.376     &  1286.3635  &      1.0000 &    0.6232 \\
        44  &  0.321E-01 &  -1.49       &    17794.    &  11.935       &        11.935     &     0.004 &    0.0045372592 &  0.0040658980 &  1.465     &   743.6684  &      1.0000 &    0.6232 \\
        45  &  0.309E-01 &  -1.51       &    17528.    &  12.178       &        12.178     &     0.005 &    0.0062184342 &  0.0053988584 & 0.8395     &   431.7279  &      1.0000 &    0.6232 \\
        46  &  0.298E-01 &  -1.53       &    17454.    &  12.403       &        12.403     &     0.007 &    0.0092269322 &  0.0077111381 & 0.5009     &   254.6085  &      1.0000 &    0.6232 \\
        47  &  0.290E-01 &  -1.54       &    17556.    &  12.559       &        12.559     &     0.009 &    0.0127924269 &  0.0103873462 & 0.3503     &   161.1637  &      1.0000 &    0.6232 \\
        48  &  0.284E-01 &  -1.55       &    17737.    &  12.669       &        12.669     &     0.012 &    0.0165275714 &  0.0131445139 & 0.2722     &   110.2394  &      1.0000 &    0.6232 \\
        49  &  0.277E-01 &  -1.56       &    18027.    &  12.791       &        12.791     &     0.015 &    0.0225027000 &  0.0174904104 & 0.2057     &    76.3695  &      1.0000 &    0.6232 \\
        50  &  0.267E-01 &  -1.57       &    18562.    &  12.956       &        12.957     &     0.022 &    0.0352738612 &  0.0266308115 & 0.1409     &    51.1543  &      1.0000 &    0.6232 \\
        51  &  0.255E-01 &  -1.59       &    19339.    &  13.140       &        13.140     &     0.034 &    0.0596374794 &  0.0438491981 & 0.9251E-01 &    33.4928  &      1.0000 &    0.6232 \\
        52  &  0.244E-01 &  -1.61       &    20058.    &  13.303       &        13.304     &     0.051 &    0.0952884943 &  0.0688903074 & 0.6365E-01 &    22.8243  &      1.0000 &    0.6232 \\
        53  &  0.235E-01 &  -1.63       &    20328.    &  13.438       &        13.438     &     0.070 &    0.1384065371 &  0.0990231779 & 0.4679E-01 &    15.8247  &      1.0000 &    0.6232 \\
        54  &  0.228E-01 &  -1.64       &    21360.    &  13.528       &        13.528     &     0.090 &    0.1843612712 &  0.1313378766 & 0.3809E-01 &    11.3048  &      1.0000 &    0.6232 \\
        55  &  0.220E-01 &  -1.66       &    21874.    &  13.633       &        13.633     &     0.117 &    0.2480407181 &  0.1765062769 & 0.2996E-01 &     8.7816  &      1.0000 &    0.6232 \\
        56  &  0.209E-01 &  -1.68       &    22854.    &  13.772       &        13.772     &     0.166 &    0.3731453797 &  0.2658365521 & 0.2180E-01 &     5.8442  &      1.0000 &    0.6232 \\
        57  &  0.194E-01 &  -1.71       &    24729.    &  13.947       &        13.948     &     0.269 &    0.6470791511 &  0.4632929887 & 0.1459E-01 &     3.6169  &      1.0000 &    0.6232 \\
        58  &  0.177E-01 &  -1.75       &    27199.    &  14.110       &        14.111     &     0.435 &    1.1312915418 &  0.8088773974 & 0.1006E-01 &     2.1696  &      1.0000 &    0.6230 \\
        59  &  0.162E-01 &  -1.79       &    29519.    &  14.224       &        14.225     &     0.628 &    1.7526133134 &  1.2371397745 & 0.7770E-02 &     1.2767  &      1.0000 &    0.6222 \\
        60  &  0.147E-01 &  -1.83       &    32188.    &  14.323       &        14.329     &     0.898 &    2.7184634806 &  1.8799146279 & 0.6203E-02 &     0.7876  &      1.0000 &    0.6188 \\
        61  &  0.127E-01 &  -1.90       &    35485.    &  14.422       &        14.440     &     1.334 &    4.3738102770 &  2.9928632321 & 0.4963E-02 &     0.5386  &      1.0000 &    0.6099 \\
        62  &  0.106E-01 &  -1.98       &    38704.    &  14.517       &        14.545     &     1.924 &    6.6095884268 &  4.5460206085 & 0.4004E-02 &     0.4105  &      1.0000 &    0.6031 \\
        63  &  0.826E-02 &  -2.08       &    42024.    &  14.622       &        14.654     &     2.757 &    9.6603255307 &  6.6991543685 & 0.3153E-02 &     0.3284  &      1.0000 &    0.6005 \\
        64  &  0.557E-02 &  -2.25       &    45795.    &  14.750       &        14.783     &     4.036 &   14.1271112197 &  9.8867578781 & 0.2361E-02 &     0.2624  &      1.0000 &    0.5995 \\
        65  &  0.313E-02 &  -2.50       &    49205.    &  14.869       &        14.902     &     5.580 &   19.2447765672 & 13.6232557658 & 0.1806E-02 &     0.2080  &      1.0000 &    0.5992 \\
        66  &  0.140E-02 &  -2.85       &    51667.    &  14.953       &        14.986     &     6.952 &   23.6095026651 & 16.8835070291 & 0.1494E-02 &     0.1746  &      1.0000 &    0.5991 \\
        67  &  0.701E-03 &  -3.15       &    52714.    &  14.987       &        15.021     &     7.587 &   25.5905904427 & 18.3799455750 & 0.1382E-02 &     0.1576  &      1.0000 &    0.5991 \\
        68  &  0.350E-03 &  -3.46       &    53248.    &  15.005       &        15.038     &     7.925 &   26.6345751524 & 19.1711773948 & 0.1328E-02 &     0.1532  &      1.0000 &    0.5991 \\
        69  &  0.175E-03 &  -3.76       &    53535.    &  15.014       &        15.047     &     8.098 &   27.1714193415 & 19.5784284504 & 0.1301E-02 &     0.1539  &      1.0000 &    0.5991 \\
        70  &   0.00     & \multicolumn{1}{c}{-0.999+100}   &    53834.    &  15.023       &        15.057     &     8.276 &   27.7224194655 & 19.9943612941 & 0.1274E-02 &     0.1539  &      1.0000 &    0.5991 \\
        \hline 
\end{longtable}
\tablefoot{A table of this kind can be retrieved from the PoWR homepage after having selected a specific model.
}
\end{landscape}
}